\documentclass[10pt,journal]{IEEEtran}
\usepackage[T1]{fontenc}
\usepackage[utf8]{inputenc}
\usepackage{authblk} 
\usepackage{amsthm} 
\usepackage{cite}
\usepackage{psfrag}
\usepackage{amssymb}
\usepackage[demo]{graphicx}
\graphicspath{ {images/} }
\usepackage{wrapfig}
\usepackage{graphicx}
\usepackage{tikz}
\usepackage{subcaption}
\usepackage{enumitem}
\usepackage{tikz} 
\usetikzlibrary{calc}
\usepackage[tbtags]{amsmath}
\usepackage{epsfig} 
\usepackage{epstopdf}
\usepackage{stfloats} 
\usepackage{float}
\usepackage{array}
\usepackage{color}
\usepackage{pgfplots}
\usepackage{textcomp} 
\usepackage{gensymb}
\usepackage{multirow}
\usepackage{url}
\usepackage{algorithm}
\usepackage{algorithmic}
\pgfplotsset{compat=newest}
\pgfplotsset{plot coordinates/math parser=false} 
\usepackage{stmaryrd}

\IEEEoverridecommandlockouts 
\newlength\tindent
\renewcommand{\indent}{\hspace*{\tindent}} 

\begin{document}
\title{Physical Layer Security for Visible Light Communication Systems: A Survey}
\author{\IEEEauthorblockN{Mohamed Amine Arfaoui},$^*$\thanks{M. A. Arfaoui and C. M. Assi are with Concordia Institute for Information Systems Engineering (CIISE), Concordia University, Montreal, Canada, e-mail:\{m\_arfaou@encs, assi@ciise\}.concordia.ca. \\ 
\indent \,\,\, M. D. Soltani, I. Tavakkolnia, M. Safari, and H. Haas are with the LiFi Research and Development Centre, Institute for Digital Communications, School of Engineering, University of Edinburgh, UK. e-mail: \{m.dehghani, i.tavakkolnia, majid.safari, h.haas\}@ed.ac.uk. \\ 
\indent \,\,\, Ali Ghrayeb is with Texas A \& M University at Qatar, Doha, Qatar, e-mail: ali.ghrayeb@qatar.tamu.edu. \\ 
\indent \,\,\, $^*$Corresponding author: M. A. Arfaoui (m\_arfaou@encs.concordia.ca). \\ 
\indent \,\,\, This work was supported by Qatar National Research Fund under NPRP Grant NPRP8-052-2-029, by FQRNT and by Concordia University. \\ 
\indent \, \, \, The statements made herein are solely the responsibility of the authors.} \IEEEauthorblockN{Mohammad Dehghani Soltani}, \IEEEauthorblockN{Iman Tavakkolnia}, \IEEEauthorblockN{Ali Ghrayeb}, \\ \IEEEauthorblockN{Chadi Assi}, \IEEEauthorblockN{Majid Safari} and \IEEEauthorblockN{Harald Haas}} 
\maketitle 
\thispagestyle{plain}
\begin{abstract} 
Due to the dramatic increase in high data rate services and in order to meet the demands of the fifth-generation (5G) networks, researchers from both academia and industry are exploring advanced transmission techniques, new network architectures and new frequency spectrum such as the visible light and the millimeter wave (mmWave) spectra. Visible light communication (VLC) particularly is an emerging technology that has been introduced as a promising solution for 5G and beyond, owing to the large unexploited spectrum, which translates to significantly high data rates. Although VLC systems are more immune against interference and less susceptible to security vulnerabilities since light does not penetrate through walls, security issues arise naturally in VLC channels due to their open and broadcasting nature, compared to fiber-optic systems. In addition, since VLC is considered to be an enabling technology for 5G, and security is one of the 5G fundamental requirements, security issues should be carefully addressed and resolved in the VLC context. On the other hand, due to the success of physical layer security (PLS) in improving the security of radio-frequency (RF) wireless networks, extending such PLS techniques to VLC systems has been of great interest. Only two survey papers on security in VLC have been published in the literature. However, a comparative and unified survey on PLS for VLC from information theoretic and signal processing point of views is still missing. This paper covers almost all aspects of PLS for VLC, including different channel models, input distributions, network configurations, precoding/signaling strategies, and secrecy capacity and information rates. Furthermore, we propose a number of timely and open research directions for PLS-VLC systems, including the application of measurement-based indoor and outdoor channel models, incorporating user mobility and device orientation into the channel model, and combining VLC and RF systems to realize the potential of such technologies.
\end{abstract} 
\begin{IEEEkeywords}
Physical layer security, visible light communication.
\end{IEEEkeywords}
\IEEEpeerreviewmaketitle 
\section{Introduction}
\subsection{Motivation} 
\begin{figure}[t] 
\centering
\includegraphics[width=1\linewidth]{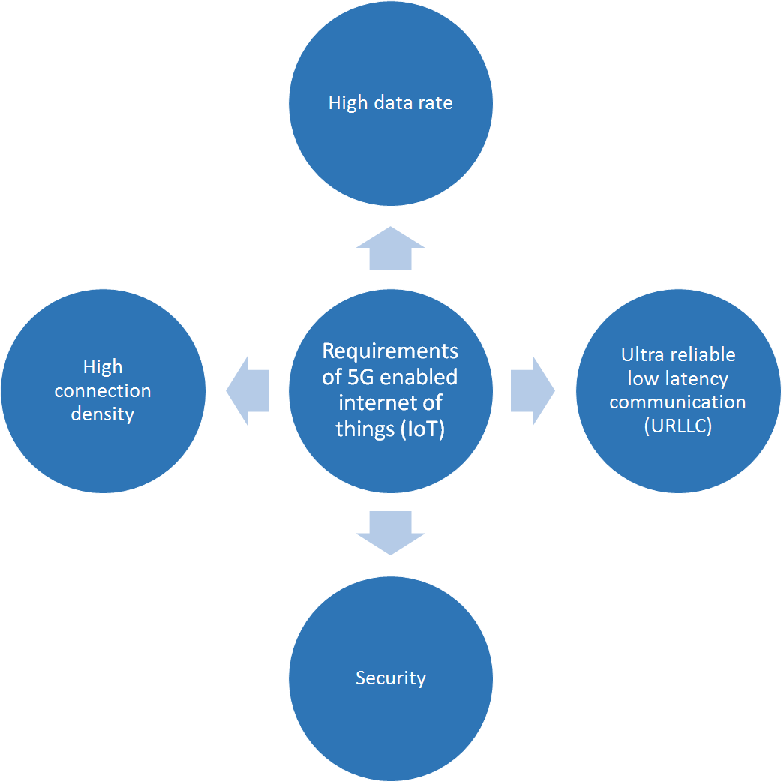}
\caption{Requirements of 5G enabled internet of things (IoT).} 
\label{fig:graph}
\end{figure}

\indent The total data traffic is expected to become about 49 exabytes per month by 2021, while  while in 2016, it was approximately 7.24 exabytes per month \cite{Intro1}. With this drastic increase, 5G networks must urgently provide high data rates, seamless connectivity, robust security and ultra-low latency communications \cite{Intro2,Intro3,Intro4}. In addition, with the emergence of the internet-of-things (IoT) networks, the number of connected devices to the internet is increasing dramatically \cite{Intro5,Intro6}. This fact implies not only a significant increase in data traffic, but also the emergence of some IoT services with crucial requirements. As shown in Fig.~\ref{fig:graph}, such requirements include high data rates, high connection density, ultra reliable low latency communication (URLLC) and security. Hence, traditional radio-frequency (RF) networks, which are already crowded, are unable to satisfy these high demands \cite{Intro7}. Network densification \cite{Intro8,Intro9} has been proposed as a solution to increase the capacity and coverage of 5G networks. However, with the continuous dramatic growth in data traffic, researchers from both industry and academia are trying to explore new network architectures, new transmission techniques and new spectra to meet these demands. One of the new communication technologies that has been proposed as an auspicious solution for 5G and beyond is visible light communication (VLC). 
\indent VLC is a promising technology that has been introduced as an auspicious solution for 5G and beyond. VLC operates in the visible light frequency spectrum and uses light for both illumination and data communication purposes simultaneously. VLC has gained significant interest due to its high data rates \cite{VLC1}. The motivation behind the interest in VLC is twofold: 1) The advantages that VLC offers when compared to RF, including the large available frequency spectrum \cite{VLC3}, high speed and robustness against interference \cite{jovicic2013visible}, and 2) the availability of low cost light emitting diodes (LEDs) \cite{Intro10}. LEDs exhibit a high electrical-to-optical conversion efficiency, long life span, low cost and high operational speed \cite{Intro11,Intro12,ref13,lin2018large}. While VLC technologies bring efficient solutions to many real-life problems, they are not intended to replace RF technologies. Rather, VLC can be viewed as a complementary technology to RF. There are applications where VLC is more efficient than RF, whereas there are other applications where the opposite is true. As such, both technologies should be exploited together to meet the expectations of 5G networks (see Fig.~\ref{fig:graph}.) and beyond, including security. Some of the noted advantages of VLC systems over RF systems is the higher security that VLC systems provide. This is basically inherited from to the fact that light does not penetrate through walls. However, security issues arise naturally in VLC systems due to their open and broadcasting nature \cite{liang2009physical}. Specifically, VLC systems could be as vulnerable as their RF counterparts when their nodes are deployed in public areas and/or when there are large windows in the coverage areas \cite{classen2015spy}. Thus, security for VLC systems is as important as it is for RF systems. \\ 
\indent Security in wireless communication systems, including 5G wireless networks, may be enhanced by introducing physical layer security (PLS) techniques \cite{yener2015wireless,yang2015safeguarding}. In fact, PLS techniques have been applied to a wide range of RF applications in an effort to improve the overall system security by complementing existing cryptography-based security techniques \cite{PLS1}. The potential of PLS stems from its ability to leverage features of the surrounding environments via sophisticated encoding techniques at the physical layer \cite{PLS3}. Indeed, PLS schemes can be applied in the same spirit to VLC systems. However, many fundamental specificities exist in the transmission protocols and modulation schemes of VLC systems that make them different from RF systems. In fact, the channels and transmitted optical signals in VLC are real and positive valued. In addition, due to the limited dynamic range of the LEDs \cite{elgala2010led}, VLC systems impose a peak-power constraint, i.e., amplitude constraint, on the channel input, which makes unbounded inputs not admissible. Note that RF signals are equally amplitude-bounded and the peak-to-average power ratio (PAPR) is a problem for RF power amplifiers. However, in VLC systems, high PAPR is used for intensity-modulation direct detection (IM/DD), which is the case for the direct current (DC) biased optical orthogonal frequency division multiplexing (DCO-OFDM) \cite{dissanayake2013comparison}. Due to these differences, PLS techniques developed for RF systems may not extend to VLC systems in a straightforward manner, which necessitates the development of new PLS schemes specific to VLC systems.
\subsection{Related Work} 
\indent Several survey papers have been reported over the past few years in the literature on VLC-related technologies  \cite{chowdhury2018comparative,VLC7,elgala2009indoor,kumar2010led,sevincer2013lightnets,VLC8,karunatilaka2015led,pathak2015visible,qiu2016channel,sindhubala2016survey,li2018optimization,zhuang2018survey,obeed2018optimizing}. However, a comprehensive and comparative study on securing VLC systems, at least from a PLS perspective, is still missing. Specifically, the authors in \cite{chowdhury2018comparative} presented the differences among various categories of optical wireless communications (OWC) technologies such as free space optical (FSO) communications, VLC and light fidelity (LiFi). Over the last few years, \cite{VLC7,elgala2009indoor,kumar2010led} were the first papers that reviewed LED-based VLC systems and their applications. In \cite{sevincer2013lightnets}, the authors brought attention to the strongest feature of LEDs, which which is their ability to provide smart lighting and data transmission simultaneously. Along the same lines, the authors of \cite{VLC8} highlighted the benefits and challenges of VLC networks, as compared to RF networks. In \cite{karunatilaka2015led}, the authors studied the methods employed for enhancing the performance of VLC, including modulation schemes and dimming control techniques, while in \cite{pathak2015visible}, the authors focused on the VLC link level transmission, the medium access techniques and the visible light sensing. On the other hand, the authors in \cite{qiu2016channel} focused on the channel models for VLC, whereas the authors in \cite{sindhubala2016survey} reviewed the optical noise sources and noise mitigation mechanisms for VLC. A study on interference reduction techniques in VLC was recently presented in \cite{li2018optimization}, where the authors reviewed and compared two different designs for VLC networks, namely, the user-centric network and network-centric designs. Positioning and localization techniques for indoor and outdoor VLC applications were reviewed in \cite{zhuang2018survey}. \\ 
\indent Recently, the authors in \cite{obeed2018optimizing} surveyed all the optimization techniques, previously reported in the literature, that aim to improve the performance of VLC systems, with emphasis on new technologies such as non-orthogonal multiple-access (NOMA), simultaneous wireless information and power transfer (SWIPT), cooperative transmission, space division multiple access (SDMA), and PLS. The same authors reviewed in \cite{obeed2018survey} the majority of the work conducted on PLS for VLC and FSO networks, and proposed several open problems to optimize and enhance the security performance of these systems. However, the coverage of PLS-VLC in these two papers is rather limited in scope and/or depth. For instance, a large number of recently published papers on this topic were not covered; some of the covered aspects, such as information theoretic security, lacked depth and breadth, owing to the relatively wide coverage scope of those papers; and several key system design challenges were not addressed. Such design issues include input signaling schemes (continuous versus discrete), transmission schemes (spatial modulation, spatial multiplexing, etc.), the geometry and parameters of VLC networks, availability of the channel state information (CSI), real-life measurement-based channel models, users mobility, devices orientation and links blockage. In this paper, however, we address all of the above-mentioned challenges where we provide an in-depth coverage of all published papers on PLS-VLC. In addition, we present a comparative study of existing techniques, and propose future research directions that incorporate several realistic system design parameters. \\  
\indent In terms of PLS for RF systems, several review articles have been reported in the literature  \cite{cao2014survey,mukherjee2014principles,atallah2015survey,chen2017survey,liu2017physical,wang2018survey}, which provided comprehensive overviews and insightful comments to understand the fundamental principles, technology status, and future trends of PLS. In \cite{cao2014survey}, security features, security vulnerabilities, and existing security solutions for long term evolution (LTE) and LTE-advanced were reviewed. In \cite{mukherjee2014principles}, the authors reviewed the secrecy performance from a PLS perspective of broadcast and multi-antenna systems, with an emphasis on the multiple-access channel (MAC), relay channels, physical-layer key generation and secure coding. The authors in \cite{atallah2015survey} reviewed different jamming techniques, employed in the literature, that aim at improving the secrecy performance of RF networks. In \cite{chen2017survey}, the authors provided a comprehensive review on various multiple-antenna techniques in PLS, with an emphasis on transmit beamforming designs for multiple-antenna nodes in point-to-point systems and heterogeneous networks. In \cite{liu2017physical}, the fundamentals and technologies of PLS were reviewed and the technologies, challenges, and solutions were also summarized from different methodological viewpoints that involve wiretap coding, multi-antenna and relay cooperation and physical-layer authentication. In \cite{wang2018survey}, security designs from optimization and signal processing viewpoints were reviewed, where the authors summarized all the PLS techniques pertaining to resource allocation, beamforming/precoding, antenna/node selection and cooperation. Although several PLS techniques were proposed in the literature for RF systems, as mentioned above, the adoption of techniques developed for RF channels can not be straightforwardly applied to VLC channels.
\subsection{Contributions and Outline}  
Most of the research done on PLS-VLC can be classified as either information theoretic, i.e., secrecy capacity, achievable secrecy rate and capacity-equivocation region; or signal processing-based, i.e., precoding, beamforming, and optimization. This paper aims at providing a unified overview of all PLS-VLC related studies that have been published so far. Among the features considered in this study are: 1) the characteristics of the VLC channel; 2) the input distribution (continuous versus discrete); 3) the architectures of the transmitter/receiver; 4) the numbers of legitimate users and unauthorized receivers; 5) the availability of channel state information (CSI); and 6) the geometry of the place of deployment and the type of signaling scheme employed. Four types of VLC systems are studied in this paper, which are the single-input-single-output (SISO), the multiple-input-single-output (MISO), the multiple-input-multiple-output (MIMO) and the hybrid RF/VLC. For each type, both cases of single active user (AU) and multiple AUs are considered and the effect of the number and CSI of the eavesdroppers (EDs) on the secrecy performance is also discussed. Furthermore, several open research problems are proposed to further advance the state-of-the-art of VLC technologies. \\
\begin{table}[t]
\caption{Table of Abbreviations}
\label{T1}
\begin{center}
\renewcommand{\arraystretch}{1.2} 
\setlength{\tabcolsep}{0.5cm} 
\begin{tabular}{| l | l |}
  \hline 
  5G & Fifth generation \\ 
  \hline
  AP & Access point \\ 
  \hline
  AU & Active user \\ 
  \hline
  CSI & Channel state information \\ 
  \hline
  DD & Direct detection \\ 
  \hline
  ED & eavesdropper \\ 
  \hline
  IM & Intensity modulation \\ 
  \hline
  LED & Light emitting diode \\ 
  \hline 
  LiFi & Light-fidelity \\
  \hline
  LoS & Line of sight \\ 
  \hline 
  MIMO & Multiple-input-multiple-output \\ 
  \hline
  MISO & Multiple-input-single-output \\ 
  \hline
  NOMA & Non orthogonal multiple access \\ 
  \hline
  PD & photo diode \\ 
  \hline
  PLS & Physical layer security \\ 
  \hline 
  PPP & Poisson point process \\ 
  \hline
  RF & Radio-frequency \\ 
  \hline 
  SISO & Single-input-single-output \\ 
  \hline
  SINR & Signal to interference plus noise ratio \\
  \hline
  SNR & Signal to noise ratio \\
  \hline
  SOP & Secrecy outage probability \\ 
  \hline
  VLC & Visible light communication \\ 
  \hline
\end{tabular} 
\end{center}
\end{table}
\indent The rest of the paper is outlined as follows. In Section II, the generalized model of VLC wiretap systems is presented. Specifically, in this section, the paper provide a brief overview about the VLC channel model adopted in the literature and discusses the constraints that must be taken into account in designing the transmission strategies and PLS schemes for VLC. In Section III, all the PLS schemes employed for the MIMO VLC system are reviewed. In Section IV, all the works related to secure MISO VLC systems from a PLS point of view are reviewed. In Section V, all the PLS techniques employed for the SISO VLC system are reviewed. In Section VI, the secrecy performance of hybrid wireless systems that combine both RF and VLC transmissions is reviewed. In Section VII, some open research directions on securing VLC systems with their associated challenges are presented. Specifically, in this section, the paper discusses several problems that have not been investigated, and proposes various ideas on how to improve the secrecy performance of real-life and practical VLC systems. In Section VIII, a summary of the paper is provided. 
\subsection{Notations and Abbreviations}
The following notations are adopted throughout the paper. Upper case bold characters denote matrices and lower case bold characters denote column vectors. $\llbracket 1,N \rrbracket$ denotes the discrete interval with bounds $1$ and $N$, $\mathbb{N}$ denotes the set of natural numbers, $\mathbb{R}^N$ denotes the set of $N$-dimensional real-valued vectors and $\mathbb{R}_+^N$ denotes the set of $N$-dimensional real-valued vectors with positive elements. $\{ \cdot \}^T$ denotes  transpose operator, $\{ \cdot \}^{-1}$ denotes matrix inversion and $| \cdot |$ denotes the absolute value of the determinant. $||\cdot||_\infty$ and $||\cdot||_2$ denote the infinity norm and the Euclidean norm, respectively. $F_\textbf{s}$ and $f_s$ denote the joint cumulative distribution function and the joint probability mass/density function of $\textbf{s}$. $\mathbb{E}( \cdot )$ denotes the expected value, $I(\cdot;\cdot)$ denotes the mutual information and $C^+ = \max(C,0)$. We use $\log [\cdot]$, without a base, to denote natural logarithms and information rates are specified in (nats/s/Hz). A list of abbreviations used in this paper is presented in Table~\ref{T1}.
\section{Generalized Model of VLC Wiretap Systems}
\label{SY1}
\begin{figure}[t]
\centering
\includegraphics[width=0.75\linewidth]{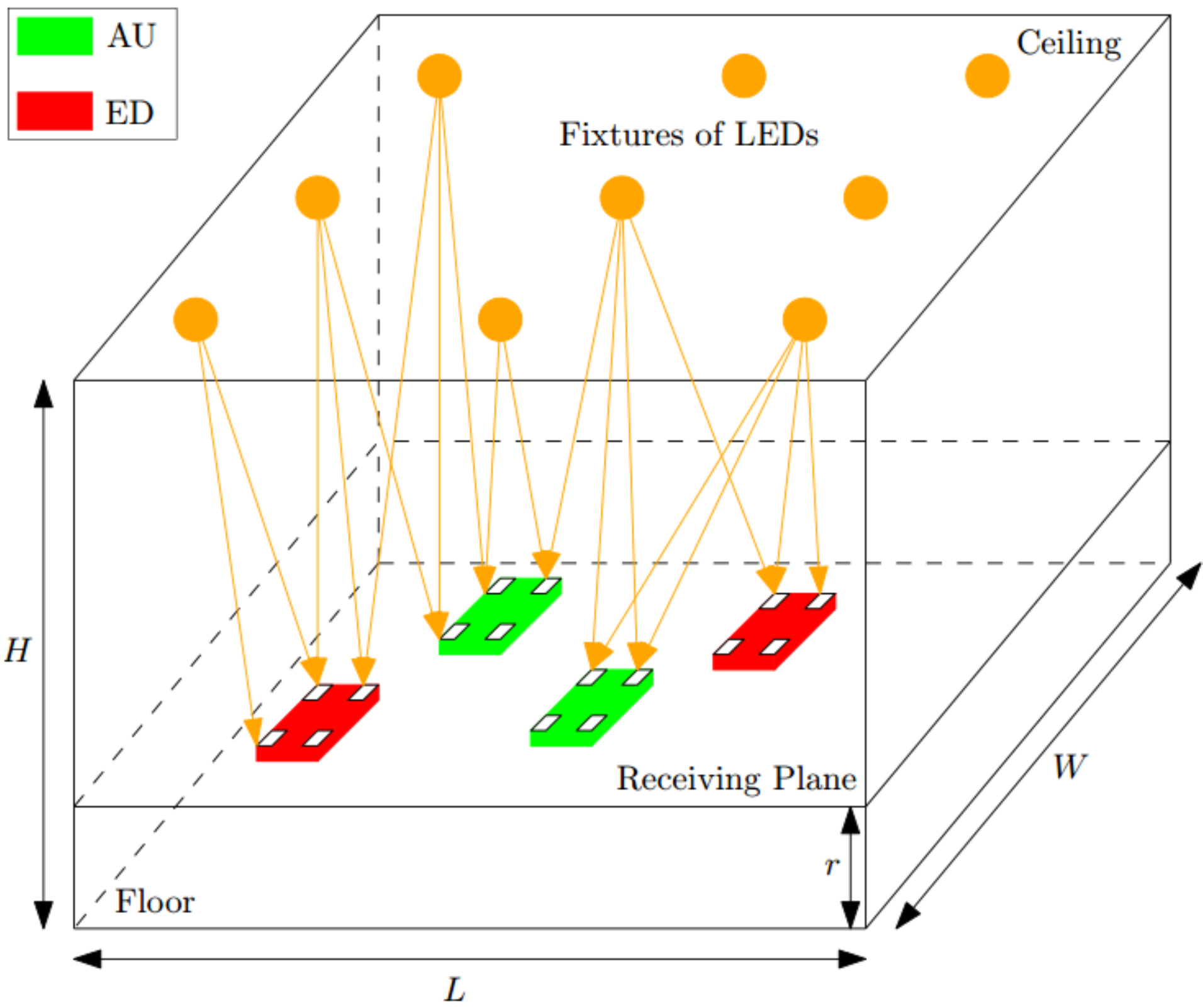} 
\caption{A MIMO VLC system consisting of a transmitter with $M = 9$ APs, multiple AUs and multiple EDs. All receivers are equipped with 4 PDs each (highlighted in white).}
\label{fig:MIMOsystem}
\end{figure}

\begin{figure}[t] 
\centering
\includegraphics[width=0.75\linewidth]{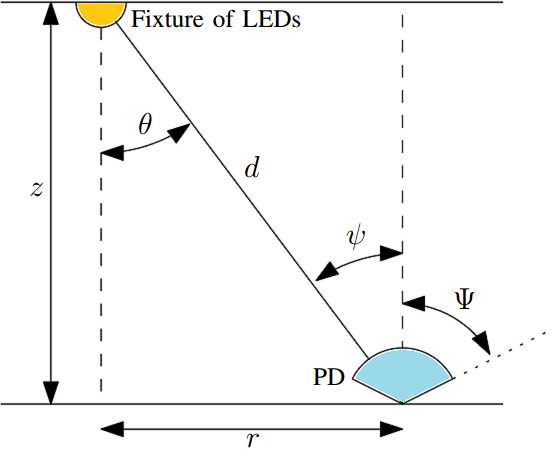} 
\caption{VLC path gain description.} 
\label{fig:Fig1}
\end{figure}

Consider the IM/DD VLC wiretap system shown in Fig. \ref{fig:MIMOsystem}, which consists of a room of size $L \times W \times H$, where $L$, $W$ and $H$ denotes the length, the width and the height of the room. The transmitter, assumed to be equipped with $M$ APs and located on the ceiling of the room, wants to transmit $N$ sets of confidential messages to $N$ AUs in the presence of $K$ EDs. For all $i \in \llbracket 1,N \rrbracket$, the $i$th AU is equipped with $N_i \in \mathbb{N}$ PDs, whereas for all $k \in \llbracket 1,K \rrbracket$, the $k$th ED is equipped with $K_j \in \mathbb{N}$ PDs. At each channel use,  the $N$ sets of confidential messages are encoded into an $M \times 1$ zero mean vector $\textbf{s}$ and then transmitted through the $M$ LED arrays. Based on this, the transmitted signal $\textbf{s}$ is expressed as
\begin{equation}
\textbf{s} = \sum_{i=1}^N \textbf{s}_i,
\end{equation}
where, for all $i \in \llbracket 1,N \rrbracket$, $\textbf{s}_i$ is the $M \times 1$ precoded signal intended to the $i$th AU. Based on the above, for $i \in \llbracket 1,N \rrbracket$ and $k \in \llbracket 1,K \rrbracket$, the received signals at the $i$th AU and the $k$th ED are expressed, respectively, as 
\begin{equation}
\begin{split}
&\textbf{y}_{i} = \textbf{H}_i \textbf{s} + \textbf{n}_{AU,i} \\ 
&\textbf{z}_{k} = \textbf{G}_k \textbf{s} + \textbf{n}_{ED,k},
\end{split}
\end{equation}
where $\textbf{H}_i$ and $\textbf{G}_k$ are the $N_i \times M$ and $K_k \times M$ channel matrices of the $i$th AU and the $k$th ED, respectively, and $\textbf{n}_{AU,i}$ and $\textbf{n}_{ED,k}$ are multivariate additive white Gaussian noise (AWGN) that are $\mathcal{N} \left(0,\sigma_A^2 \textbf{I}_{N_i} \right)$ and $\mathcal{N} \left(0,\sigma_E^2 \textbf{I}_{K_k} \right)$ distributed, respectively. It is important to highlight that, since the legitimate receivers are usually active, the CSI of each AU is available at the transmitter through a limited-feedback mechanism \cite{dehghani2015limited,soltani2017throughput,soltani2018bidirectional}. However, this is not the case for the EDs. In fact, the CSI of the EDs are not available at the transmitter in general, especially if they are passive or not registered in the network. \\ 
\indent The system model presented above refers to the generalized MIMO VLC wiretap system \cite{khisti2010secure2} and it includes all varieties of VLC wiretap systems for any number of APs $M$, any number of AUs $N$ and any number of EDs $K$, as well as any number of PDs at each receiver. When all receivers are equipped with a single PD, the system is scaled down to the MISO VLC wiretap system. Additionally, when the transmitter is equipped with one AP, i.e., $M=1$, the system corresponds to the SISO VLC wiretap system. The secrecy performance and the suitable PLS techniques of VLC wiretap systems depend mainly on the parameters $M$, $N$ and $K$, as well as the number of PDs and the availability of the CSI of each receiver. \\ 
\indent It is important to mention that almost all work on secure VLC systems have only considered the LoS component $h_{\text{LoS}}$ presented in (2) in their adopted VLC channel models and ignored the NLoS component $h_{\text{NLoS}}$. This is mainly due to two reasons: 1) no closed-form expression for the NLoS component $h_{\text{NLoS}}$ is readily available in the literature, and 2) the optical power received from signals reflected more than once is negligible compared to the LoS component, especially if the receiver is far away from the walls or is located close to the cell center \cite{komine2004fundamental}. Thus, unless otherwise stated, only the LoS component is considered in the analysis presented in this paper. In this case, assuming that the considered LEDs have a Lambertian emission pattern, the LoS component $h_{\text{LoS}}$ of the channel gain between one fixture of LEDs and one PD is expressed as \cite{Intro13,gfeller1979wireless}
\begin{equation}
h_{\text{LoS}} = \eta R_p T \frac{(m+1)}{2 \pi} \frac{n_c^2 A_{g}}{\sin(\Psi)^2}  \cos^m(\theta) \frac{ \cos(\psi)}{d^2} \text{rect} \left( \frac{\psi}{\Psi} \right), 
\end{equation}
where $m= \frac{-\log (2)}{\log(\cos(\theta_{1/2}))}$ is the order of the Lambertian emission, such that $\theta_{1/2}$ represents the half-power semi-angle of the LED, $A_{g}$ is the geometric area of the PD, $n_c$ is the refractive index of the receiver's optical concentrator, and as shown in Fig \ref{fig:Fig1}, $\Psi$ is the field of view (FoV) of the receiver's PD, $d = \sqrt{r^2+z^2}$ is the Euclidean distance between the LED and the PD, such that $r$ and $z$ are the horizontal and vertical distances between the LED and the PD, $\theta \in [0, \theta_{1/2}]$ is the radiation angle and $\psi \in [0, \pi]$ is the incidence angle.  \\
Typical LEDs suffer from nonlinear distortion and clipping effects, which imposes additional operating constraints on the emitted optical power \cite{mesleh2012led,elgala2009study,elgala2009non}. Due to this, the transmitted signal $\textbf{s}$ should satisfy a peak-power constraint, i.e., an amplitude constraint, that is expressed as
\begin{equation}
||\textbf{s}||_\infty \leq A, 
\end{equation}
where $A$ is the maximum allowed signal amplitude at the input of each fixture of LEDs. The peak-power constraint implies that the information bearing signal $\textbf{s}$ is bounded. Consequently, from an information-theoretic perspective, the capacity achieving input distribution can not be Gaussian, since Gaussian inputs are not admissible for such constraints \cite{lapidoth2009capacity}. 
However, it is shown in \cite{dimitrov2013information} that the channel can be modelled in the \textit{electrical} domain as an AWGN channel with an average power constraint, and this falls within the Shannon framework which determines the upper bound on the achievable data rate.
\indent In the following sections, we review all the works related to secure VLC systems as well as all the associated PLS techniques reported in the literature. For ease of presentation, we group those techniques according to what system model they have been developed for, e.g., MIMO, MISO and SISO.   
\section{The MIMO VLC Wiretap System} 
\subsection{Single ED}
\begin{figure}[t]
	\centering
	\includegraphics[width=1\linewidth]{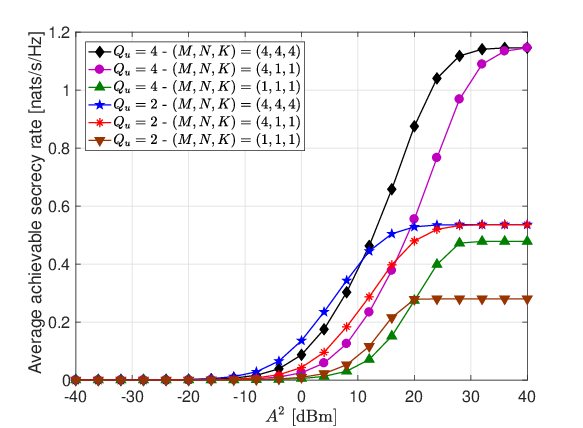}
	\caption{Average achievable secrecy rate $R_{s,1}^+$ versus the square of the amplitude constraint $A^2$, for the system configurations $(M,N,K)=(4,4,4)$, $(4,1,1)$ and $(1,1,1)$ and for the number of mass points $Q_u=2$ and $Q_u=4$.}
	\label{fig:Sim1}
\end{figure}

For the special case when $N=1$ AU and $K=1$ ED, the secrecy capacity of the MIMO VLC wiretap channel in (2) is given by \cite{aghdam2018overview,dytso2019amplitude}
\begin{equation}
\begin{aligned}
 C_s = \,\, &\underset{F_{\textbf{s}}}{\max} \, \, \left[ \mathrm{I} \left(\textbf{s};\textbf{y}_1 \right) - \mathrm{I} \left(\textbf{s};\textbf{z}_1 \right) \right]^+ \\ 
 &\text{s.t} \int_{\mathcal{S}} \mathrm{d} F_{\textbf{s}}(\textbf{x}) = 1,
\end{aligned}
\end{equation}
where $\mathcal{S} = \left[-A,A \right]^M$. In general, neither the secrecy capacity $C_s$ nor the optimal probability distribution that achieves $C_s$, i.e., the one that solve the optimization problem in (5), have been determined in closed-form expression or at least characterized in the literature, except some special cases that will be discussed later in the paper. Due to this, based on the system configuration and the type of the input signaling scheme, several upper and lower bounds were derived in the literature in order to characterize the secrecy performance of the MIMO VLC wiretap channel. In this context, it should be noted that conceptually there are two approaches for transmission design of MIMO wiretap channels. One is to perform precoding under fixed input probability distribution and system constraints. The other is to optimize the input distributions that can achieve secrecy rates as close as possible to the secrecy capacity \cite{wu2018survey1}.\\
\indent Although security for VLC systems was extensively studied in the literature, only a few papers investigated the secrecy performance of MIMO VLC systems \cite{arfaoui2017mimo,arfaoui2018TWC2,le2014secured,chen2017physical}. For the case of $N=1$ AU and $K=1$ ED, the authors in \cite{arfaoui2017mimo} derived an achievable secrecy rate for the system using continuous log-concave distributions. Afterwards, an iterative algorithm based on the convex-concave procedure (CCP) was proposed to jointly obtain the best covariance matrix and signaling scheme. From the information-theoretic point of view, even if continuous input signaling schemes achieve the secrecy capacity of MIMO system, continuous transmit signals are rarely used in practical communication systems. This is mainly due to the fact that, if the probability density function (pdf) of a  transmit signal is continuous, the task of signal detection at the receiver will be significantly complicated. Therefore, in practice, transmit signals are discrete signals drawn from finite discrete constellations, such as pulse amplitude modulation (PAM) in the context of optical wireless communication. Due to this fact, the authors in \cite{arfaoui2018TWC2} derived an achievable secrecy rate for the same system using discrete distributions with finite support sets. This achievable secrecy rate is given by $R_{s,1}^+$, where 
\begin{equation}
\begin{split}
R_{s,1} &= \frac{1}{2} \log \left[ \det \left(\textbf{B} \right) \right] - \log \left[ \sum_{i=1}^{Q} \sum_{j=1}^{Q} p_i p_j \exp \left(\frac{d_{i,j}}{2\sigma_A^2}\right) \right] \\ 
&- \frac{1}{2} \log \left[\det \left(\textbf{I}_{K_1} + \frac{\textbf{G}_1 \textbf{K}_{\textbf{s}} \textbf{G}_1^T}{\sigma_E^2}  \right) \right],
\end{split}
\end{equation}
in which $\textbf{K}_{\textbf{s}}$ represents the covariance matrix of the transmitted signal $\textbf{s}$,  $\textbf{B} = 2 \textbf{I}_{N_1} - \left(\textbf{I}_{N_1} + \frac{\textbf{H}_1 \textbf{K}_{\textbf{s}} \textbf{H}_1^T}{\sigma_A^2} \right)^{-1}$ and, for all $i,j \in \llbracket 1,Q \rrbracket$,
\begin{equation}
\begin{split}
d_{i,j} &= \left(\textbf{q}_i + \textbf{q}_j \right)^T \textbf{H}_1^T \textbf{B}^{-1}\textbf{H}_1 \left(\textbf{q}_i + \textbf{q}_j \right) \\ 
& - \left| \left| \textbf{H}_1 \textbf{q}_i \right| \right|_2^2 - \left| \left| \textbf{H}_1  \textbf{q}_j \right| \right|_2^2,
\end{split} 
\end{equation}
such that $Q$ represents the number of mass points of $\textbf{s}$, $\left\{ \textbf{q}_i \left| i\in \llbracket 1,Q \rrbracket \right. \right\}$ is its set of mass points and $\left\{ p_i \left| i\in \llbracket 1,Q \rrbracket \right. \right\}$ is its set of mass probabilities. The core idea behind the achievable secrecy rate $R_{s,1}^+$ is based on the relation embedded on the Kullback–Leibler (KL) divergence, also known as the relative entropy, between continuous and discrete input distributions that have the same covariance matrix \cite{cover2006elements}. In addition, the achievable secrecy rate $R_{s,1}^+$ is valid for all possible configurations of the considered MIMO VLC wiretap channel, i.e., MIMO, MISO and SISO VLC wiretap channels. \\
\indent Motivated by the result in (6), the authors in \cite{arfaoui2018TWC2} investigated the case where the transmitter aims to send a zero-mean vector $\textbf{u}$ containing $L$ confidential messages, where $1\leq L \leq M$, to the AU in the presence of an ED. Therefore, the transmitted signal is expressed as $\textbf{s} = \textbf{s}_1 = \textbf{W}\textbf{u}$, where $\textbf{W}$ is the associated $M \times L$ precoding matrix. In this case, $\textbf{u}$ and $\textbf{W}$ were assumed to satisfy $||\textbf{u}||_\infty \leq A$ and $||\textbf{W}||_\infty \leq 1$ in order to fulfill the amplitude constraint in (4). Moreover, the authors in \cite{arfaoui2018TWC2} assumed that the confidential messages are independent and identically distributed (i.i.d) according to a generic scalar random variable $u$ that follows a truncated discrete generalized normal (TDGN) distribution within $[-A,A]$. Moreover, the ED was assumed to be randomly located within the coverage area and a precoding scheme based on the generalized singular value decomposition (GSVD) of the channel matrices was proposed in order to enhance the secrecy performance of the system. \\
\indent As an illustration, Fig.~\ref{fig:Sim1} presents the average achievable secrecy rate $R_{s,1}^+$, obtained through $10^5$ independent Monte-Carlo trials on the location of the AU and the ED, versus the square of the amplitude constraint $A^2$ in [dBm], for the system configurations $(M,N,K)=(4,4,4)$, $(4,1,1)$ and $(1,1,1)$ and for the number of mass points $Q_u=2$ and $Q_u=4$ of the random scalar variable $u$. The number of confidential messages is fixed to $L=1$ and the precoding matrix $\textbf{W}$ was optimized using the brute force (BF) search methods. The noise variance at the two receivers is fixed to $\sigma_A^2 = \sigma_E^2 = -68.93$ [dBm]. Fig.~\ref{fig:Sim1} shows that increasing the number of light sources of the transmitter and/or increasing the number of PDs at the receivers will increase the achievable secrecy rate $R_{s,1}^+$. In other words, it shows that increasing the system diversity will improve the secrecy performance of the system. In addition, Fig.~\ref{fig:Sim1} shows that the best number of mass points $Q_u$ is a function of the operating range of the amplitude constraint $A$. \\
\indent On the other hand, an upper bound on the secrecy capacity $C_s$ was derived in \cite{arfaoui2017mimo,arfaoui2018TWC2}. The key idea was converting the amplitude constraint into an average power constraint as $||\textbf{s}||_\infty \leq A  \implies \text{Trace}(\textbf{K}_s) = \mathbb{E} \left( \textbf{s}^T\textbf{s} \right) \leq M A^2$, which is the trace constraint on the input covariance. The derived upper bound is given by 
\begin{equation} 
\begin{split}
U_B = \, \, &\underset{\textbf{K}_{\textbf{s}} \geq 0}{\max} \, \,  \frac{1}{2} \log \left[ \frac{\det\left( \textbf{I}_M + \frac{1}{\sigma_A^2} \textbf{H}_1^T \textbf{H}_1 \textbf{K}_\textbf{s} \right)}{\det\left(\textbf{I}_M + \frac{1}{\sigma_E^2} \textbf{G}_1^T \textbf{G}_1 \textbf{K}_\textbf{s} \right)} \right] \\
 &\text{s.t.} \, \, \, \text{Trace}(\textbf{K}_\textbf{s}) \leq MA^2.
\end{split}
\end{equation} 
It was stated in \cite{loyka2016optimal} that the optimal covariance matrix, solution of the optimization problem in (8), is expressed through the active eigenvectors, i.e., the ones associated to positive eigenvalues, of the matrix $\frac{\textbf{H}_1^T \textbf{H}_1}{\sigma_A^2} - \frac{\textbf{G}_1^T \textbf{G}_1}{\sigma_E^2}$. Based on this, the closed form expression of the upper bound (8) was derived for the special case where the number of PDs at the receivers is one, i.e., the SISO and MISO VLC wiretap systems. However, for the general case, iterative approaches were employed in \cite{arfaoui2017mimo,arfaoui2018TWC2} in order to solve the optimization problem in (8).
\subsection{Multiple EDs}
\indent For a MIMO VLC system with $N=1$ AU and multiple EDs, the authors in \cite{le2014secured} established a secure communication mechanism by minimizing the bit error rate (BER) in a protected zone and maximizing it everywhere else. On the other hand, for a MIMO VLC system with multiple AUs and multiple EDs, the authors of \cite{chen2017physical} improved the secrecy performance by using a continuous signaling scheme and applying angle diversity transmitters that are capable of transmitting data in narrow beams and effectively minimizing the leakage of information. By comparing different types of optical network deployments, they concluded that the hexagonal deployment is the best in terms of secure communications, whereas the Poisson point process (PPP) deployment is the worst.
\section{The MISO VLC Wiretap System}
In this section, each receiver is equipped with a single PD. The transmitter intends to transmit $N$ sets of confidential messages to $N$ spatially dispersed AUs in the presence of $K$ EDs. As such, for all $i \in \llbracket 1,N \rrbracket$ and $k \in \llbracket 1,K \rrbracket$, the received signals at the $i$th AU and the $k$th ED are expressed, respectively, as 
\begin{equation}
\begin{split}
&y_i = \textbf{h}_{i}^T \textbf{s} + n_{A,i} \\ 
&z_k = \textbf{g}_{k}^T \textbf{s} + n_{E,k},
\end{split}
\end{equation}
where $\textbf{h}_i, \textbf{g}_k \in \mathbb{R}_+^M$ are the $M \times 1$ channel gain vectors of the $i$th AU and the $k$th ED, respectively, and $n_{A,i}$ and $n_{E,k}$ are AWGN samples that are $\mathcal{N} \left(0, \sigma_A^2 \right)$ and $\mathcal{N} \left(0, \sigma_E^2 \right)$ distributed, respectively. In the following subsections, we review all the works reported in the literature on secure MISO VLC systems for the two cases when the transmitter communicates with a single AU or with multiple AUs. 
\subsection{Single AU} 
\begin{figure}[t]
	\centering
	\includegraphics[width=1\linewidth]{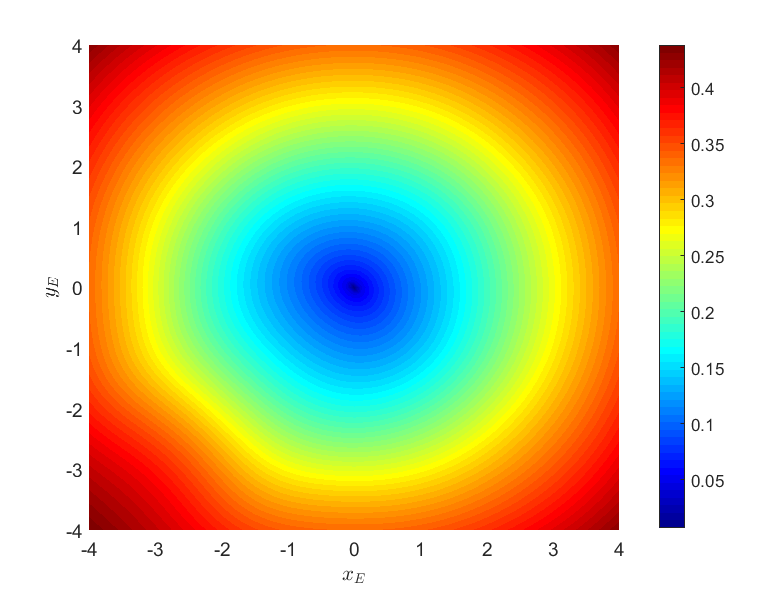}
	\caption{Optimal fraction of amplitude $\alpha$ allocated to the information bearing signal versus the location of the ED within a room of size $8$m$\times8$m$\times4$m. The transmitter is equipped with $M=16$ APs and the AU is located at the center of the room.}
	\label{fig:Sim2}
\end{figure}

For a single AU, a large body of work in the literature proposed various PLS techniques that aim to provide secure communications, for both cases of single ED \cite{mostafa2014physical,mostafa2015physical,mostafa2016optimal,arfaoui2016input,mostafa2014securing,zaid2015improved,chosecuring,chosecuringJ,arfaoui2016secrecy,arfaoui2017discrete,shen2016secrecy,arfaoui2018TCOM,wang2018optical,wang2018secrecy,li2018secrecy} and multiple EDs  \cite{cho2017secrecy,cho2018securing,arfaoui2018enhancing,chow2015secure,arfaoui2018discrete}.  In the following, both cases of single and multiple EDs will be studied, respectively. \\ 
\indent \textbf{Single ED.}	In this case, it was shown in \cite{rezki2017secret} that the optimal joint probability distribution $f_\textbf{s}$ of the transmitted signal $\textbf{s}$ that achieves the secrecy capacity of the system is unique, symmetric and discrete with a finite support set. However, neither the secrecy capacity nor the optimal input probability distribution were derived in closed-form and, thus, it remains an open problem. Therefore, both continuous \cite{mostafa2014physical,mostafa2015physical,mostafa2016optimal,arfaoui2016input,mostafa2014securing,zaid2015improved,chosecuring,chosecuringJ,arfaoui2016secrecy} and discrete \cite{arfaoui2017discrete,shen2016secrecy,arfaoui2018TCOM,wang2018optical,wang2018secrecy,li2018secrecy} signaling schemes were employed, in order to characterize the secrecy capacity of the system. In addition, beamforming and artificial noise based beamforming are the widely used PLS techniques for securing MISO VLC systems. \\ 
\indent For the case of continuous signaling schemes, beamforming was employed in \cite{mostafa2014physical,mostafa2015physical,mostafa2016optimal,arfaoui2016input}. With beamforming, the transmitted signal is expressed as $\textbf{s} = \textbf{v} u$, where $u$ denotes the confidential message intended to the AU and $\textbf{v}$ denotes its associated beamforming vector. In \cite{mostafa2014physical,mostafa2015physical,mostafa2016optimal}, the pdf of $u$ was assumed to be uniform within $[-A,A]$ and the authors investigated the performance of beamforming for both scenarios, namely, when the ED's CSI is available at the transmitter or not. In the former scenario, zero-forcing (ZF) beamforming, also known as null steering, was employed in order to force the ED's reception to zero, i.e., $\textbf{g}_1^T\textbf{v}=0$. However, in the latter case, the ED was assumed to be located within a predefined area known to the transmitter and robust beamforming was employed in order to maximize the worst case achievable secrecy rate. In \cite{arfaoui2016input}, the authors used the truncated generalized normal (TGN) distribution as a probability distribution for $u$ and derived closed-form expression for the optimal beamformer that maximizes the achievable secrecy rate of the system. In addition, they optimized over the TGN parameters to further enhance the secrecy performance of the system. \\ 
\indent Artificial noise based beamforming was adopted in \cite{mostafa2014securing,zaid2015improved,chosecuring,chosecuringJ,arfaoui2016secrecy}. Specifically, in \cite{mostafa2014securing,zaid2015improved}, the proposed schemes consist of transmitting the information bearing signal from one AP and jamming signals from the remaining APs. Moreover, the information bearing and jamming signals were assumed to follow uniform distributions in \cite{mostafa2014securing} and truncated Gaussian distributions in \cite{zaid2015improved}. In \cite{chosecuring,chosecuringJ}, the authors proposed an artificial noise based beamforming scheme where, based on the available CSI of the AU and the ED, each transmitter's AP can choose to convey data or jamming signals. In \cite{arfaoui2016secrecy}, the information bearing signal and the jamming signals were transmitted jointly through all APs. Specifically, the proposed scheme in \cite{arfaoui2016secrecy} consists of simultaneously transmitting a scalar information signal $u$ and a scalar jamming signal $x$ through two different beamforming vectors $\textbf{v}$ and $\textbf{w}$, respectively. In this case, the transmitted signal is given by $\textbf{s} = \textbf{v}u + \textbf{w}x$. In addition, both $u$ and $x$ were assumed to follow a TGN distribution and the effect of the jamming signal was forced to be canceled at the AU, i.e., $\textbf{h}_1^T\textbf{w}=0$. Moreover, $u$, $x$, $\textbf{v}$ and $\textbf{w}$ were assumed to satisfy $|u| \leq \frac{A}{2}$, $|x| \leq \frac{A}{2}$, $||\textbf{v}||_\infty \leq 1$ and $||\textbf{w}||_\infty \leq 1$ in order to satisfy the the amplitude constraint $||\textbf{s}||_\infty \leq 1$ imposed on the transmitted signal $\textbf{s}$. Note that in \cite{zaid2015improved,arfaoui2016secrecy}, both cases of perfect and imperfect ED's CSI were considered, where the derived achievable secrecy rates were maximized with respect to the corresponding beamforming vectors in the former case, and robust beamforming employed in the latter case. \\ 
\indent For the case of discrete signaling schemes, beamforming was employed in \cite{arfaoui2017discrete}, where the authors tried to solve the secrecy capacity numerically since no closed-form expression of the secrecy capacity has been derived yet. In \cite{shen2016secrecy,arfaoui2018TCOM,wang2018optical}, artificial noise aided precoding was employed. Specifically, in \cite{shen2016secrecy}, the adopted secrecy performance measure is the signal to interference plus noise  ratio (SINR) and the proposed scheme consists of transmitting a scalar information bearing signal through beamforming simultaneously with different jamming signals that are linearly precoded into a multi-dimensional jamming vector, i.e., $\textbf{s} = \textbf{v} u + \textbf{W} \textbf{x}$, where $\textbf{x}$ is a vector of the jamming signals and $\textbf{W}$ is its associated precoding matrix. \\ 
\indent In the same context, the authors in \cite{arfaoui2018TCOM} recently employed the same artificial noise based beamforming scheme adopted in \cite{arfaoui2016secrecy}. However, differing from \cite{arfaoui2016secrecy}, both the information bearing signal $u$ and the jamming signal $x$ were drawn from a TDGN distribution. In addition, the ED was assumed to be randomly located in the coverage area and the authors used a stochastic geometry model to evaluate the statistics of the ED's SINR and the average achievable secrecy rate of the system. Moreover, $u$, $x$, $\textbf{w}_1$ and $\textbf{w}_2$ were assumed to satisfy $|u| \leq \alpha A$, $|x| \leq (1-\alpha)A$, $||\textbf{v}||_\infty \leq 1$ and $||\textbf{w}||_\infty \leq 1$ in order to satisfy the amplitude constraint in (4), where $\alpha \in ]0,1]$. As an illustration, consider a room of size $8$m$\times8$m$\times4$m. The transmitter is equipped with $16$ APs located on the ceiling of the room and at the positions $\left(x,y \right) \in \left\{1,3,5,7 \right\} \times \left\{1,3,5,7 \right\}$. The AU is located at the center of the room and the ED is randomly located within the room. Since the transmitter has no information about the CSI of the ED, the best transmission strategy to adopt at the transmitter is to send the information bearing signal $u$ into the direction of the AU, i.e., $\textbf{v} = \frac{\textbf{h}_1}{||\textbf{h}_1||_\infty}$, and the jamming signal into the orthogonal direction of the the AU, i.e., $\textbf{w} \in \text{Null} \left( \textbf{h}_1 \right)$. Based on this scheme, Fig.~\ref{fig:Sim2} presents the optimal fraction of amplitude $\alpha$ allocated to the information bearing signal, i.e., the one that maximizes the secrecy performance of the system, versus the location of the ED where $\frac{A^2}{\sigma_A^2} = \frac{A^2}{\sigma_E^2} = 98.83$ dB. This figure shows that when the ED is close to the AU, the amplitude fraction $\alpha$ decreases. Meaning, the transmitter have to allocate more amplitude, and power equivalently, to the jamming signals in order to degrade the reception of the ED. \\ 
\indent The potential of spatial modulation (SM) to enhance the secrecy performance of MISO VLC wiretap systems was investigated in \cite{wang2018optical,li2018secrecy,wang2018secrecy}. Specifically, in \cite{wang2018optical}, artificial noise based beamforming was proposed, where SM was used to transmit the information bearing signal and truncated Gaussian was adopted for the jamming signal. In \cite{li2018secrecy}, the authors studied the secrecy performance of the MISO VLC system employing SM, termed as the MISO SM-VLC. The authors obtained a lower bound and an accurate closed-form expression for the approximate achievable secrecy rate of the generalized space shift keying (GSSK) VLC system and also derived closed-form expressions for the pairwise error probability and bit error rate of the system. Similar results were obtained in \cite{wang2018secrecy}, where the authors proposed a LED pattern selection algorithm in order to enhance the secrecy performance of the GSSK VLC system.  \\
\indent \textbf{Multiple EDs.} For the case of multiple EDs, both continuous \cite{cho2017secrecy,cho2018securing,arfaoui2018enhancing,chow2015secure} and discrete \cite{arfaoui2018discrete} signaling schemes were employed. However, only beamforming was employed to secure this class of VLC systems. In \cite{cho2017secrecy,cho2018securing}, the authors proposed a LED selection scheme to improve the secrecy outage probability (SOP) of the system. The SOP represents the probability that the instantaneous secrecy capacity $C_s$ falls below a target secrecy rate $R_{th}$ \cite{pinto2009wireless}, i.e., $\mathbb{P} \left( C_s \leq R_{th} \right)$. However, due to the amplitude constraint imposed on the transmitted signal, the secrecy capacity is not readily available and a modified SOP was employed as a secrecy performance measure instead. In addition, the authors in \cite{cho2017secrecy,cho2018securing} assumed that the EDs are not colluding. In this case, only the ED with the strongest SNR is considered as a potential threat. Moreover, the EDs were assumed to be randomly located in the coverage area and thus their CSI is unknown to the transmitter. In such a case, stochastic geometry approaches can be used to characterize the system secrecy performance. These approaches are based on proposing stochastic geometric models for the channel gains of the EDs \cite{haenggi2008secrecy,pinto2008physical}. Stochastic geometry is a powerful tool for dealing with spatial uncertainty \cite{baccelli2010stochastic,haenggi2012stochastic}. In fact, when the EDs are distributed randomly in the considered area, stochastic geometry can be used to determine the statistics of their individual CSI and, thus, it can provide closed-form expressions for typical secrecy performance measures. In \cite{cho2017secrecy,cho2018securing}, the number of EDs was modeled using the PPP model and the locations of each ED were assumed to be uniform within the coverage area. Based on this, the authors employed a stochastic geometry model to derive a closed-form expression of the SOP and proposed a beamforming scheme that enhances the secrecy performance. \\ 
\indent The same problem but with colluding and randomly located EDs was considered in \cite{arfaoui2018enhancing}. The authors employed beamforming as a transmission strategy and used stochastic geometry to derive an approximate expression of the average achievable secrecy rate. Then, the resulting problem of optimal beamforming that maximizes the average achievable secrecy rate was solved. In \cite{chow2015secure}, an experimental artificial noise based precoding scheme was proposed. The experimental setup consists of eight arrays of LEDs, where four arrays are used for data transmission and four arrays are used for artificial noise transmission. Experimental and simulation results show that the proposed scheme highly outperforms the scheme that does not inject artificial noise. In \cite{arfaoui2018discrete}, the EDs were assumed to be colluding and randomly located similar to \cite{arfaoui2018enhancing}. However, differing from \cite{arfaoui2018enhancing}, only discrete input signaling schemes were employed and the authors derived an average achievable secrecy rate using stochastic geometry and proposed a beamforming solution that maximizes the resulting average achievable secrecy rate. 
\subsection{Multiple AUs}
\label{MultipleAUs}
Several studies in the literature have proposed transmission strategies and precoding designs to secure MU-MISO VLC broadcast channels  \cite{mostafa2017linear,pham2015max,pham2017multi,arfaoui2018TWC1,pham2016secrecy,pham2017secrecy,pham2018artificial,pham2018artificialJ,yin2018physical}. It was assumed in \cite{mostafa2017linear,pham2015max,pham2017multi,arfaoui2018TWC1} that there were no EDs, but the AUs were treated as potential EDs when a message was not intended for them, i.e., when the messages were confidential. In \cite{pham2016secrecy,pham2017secrecy,pham2018artificial,pham2018artificialJ,yin2018physical}, however, external EDs were assumed to coexist with the AUs, while the AUs were continually treated as potential EDs. \\ 
\indent Although it was assumed in \cite{mostafa2017linear,pham2015max,pham2017multi,arfaoui2018TWC1} that there is no external ED in the vicinity of the AUs, the transmitted messages to the AUs were assumed to be confidential, such that each AU is supposed to receive and decode only its own message, i.e., users remain ignorant about messages that are not intended to them. In other words, the transmitter has to communicate each message to its intended AU while keeping each AU unaware of the other messages. In \cite{mostafa2017linear}, the secrecy performance of a two-user MISO broadcast channel with confidential messages was considered, where a per-antenna amplitude constraint and a per-antenna power constraint were assumed and the authors investigated the problem of optimal linear precoding schemes. ZF precoding in conjunction with continuous uniform signaling scheme was employed in \cite{pham2015max,pham2017multi} to cancel information leakage between users. The same problem was investigated in \cite{arfaoui2018TWC1}, where the transmitter was assumed to communicate with multiple spatially dispersed AUs. Linear precoding schemes were developed to enhance the secrecy performance of the system. In this context, typical secrecy performance measures, such as the max-min fairness, the harmonic mean, the proportional fairness and the weighted fairness, were investigated. These secrecy performance measures are defined as follows. Let $S$ be the objective function that represents the secrecy performance measure of interest and for all $i \in \llbracket 1,N \rrbracket$, let $R_{i}$ be the achievable secrecy rate of the $i$th AU. Thus, the aforementioned secrecy performance measures are defined as \cite{luo2008dynamic}:
\begin{enumerate}[label=\roman*)]
\item Max-min fairness: $S = \underset{1 \leq i \leq N}{\min} \, R_{s,i}$.
\item Harmonic mean: $S = K \left(\sum_{i=1}^N R_{i}^{-1}\right)^{-1}$.
\item Proportional fairness: $S = \left(\prod_{i=1}^N  R_{i} \right)^{\frac{1}{K}}$.
\item Weighted fairness: $S = \sum_{i=1}^N \alpha_i R_{i}$, where $ \left(\alpha_i \right)_{1 \leq i \leq N} \in \mathbb{R}^+$,
\end{enumerate}
with an increasing order of achievable secrecy sum-rate and a decreasing order of user fairness \cite{luo2008dynamic}. \\ 
\indent The problem of secure MU-MISO brodcast channels with multiple EDs was investigated in \cite{pham2016secrecy,pham2017secrecy,pham2018artificial,pham2018artificialJ,yin2018physical}. For the case when there is only one ED, i.e., $K=1$, the authors in \cite{pham2016secrecy,pham2017secrecy} studied the potential of employing ZF precoding and artificial noise schemes in providing secure VLC communications. In the same case, an artificial noise based precoding scheme was proposed in \cite{pham2018artificial,pham2018artificialJ}. The proposed scheme was designed to solve the max-min fairness SINR problem among AUs in two different scenarios: known and unknown ED’s CSI at the transmitter. In the former case, the proposed scheme was designed in such a way that it kept the SINR of the ED below a predefined threshold, whereas in the latter case, the traditional null-space artificial noise scheme was employed, which consists of sending jamming signals in all the orthogonal directions of the AUs. For the case when $K>1$, the authors in \cite{yin2018physical} proposed a three-dimensional network model, where the VLC APs were modeled by a two-dimensional homogeneous PPP in the ceiling and the locations of of both the AUs and the EDs were modeled by another independent two-dimensional homogeneous PPP. In this case, for VLC networks with and without APs cooperation, the secrecy performance of the system was evaluated using the SOP and the ergodic secrecy rate. 
\section{The SISO VLC Wiretap System}
\subsection{System Model}
In this section, we consider the SISO VLC case where the transmitter is equipped with a single AP. The transmitter intends to transmit $N \in \mathbb{N}$ confidential messages to $N$ AUs, each equipped with a single PD, in the presence of a set of $K$ spatially dispersed EDs, each equipped with a single PD. Let $\left\{u_i | i \in \llbracket 1, N \rrbracket \right\}$ be the set of confidential messages intended to the AUs, where for all $i  \in \llbracket 1, N \rrbracket$, $u_i$ is the intended confidential message to the $i$th AU. The confidential messages are encoded via superposition coding (SC) into one scalar signal $s = \sum_{i=1}^N u_l$, that should satisfy the amplitude constraint in (4), i.e., $|s| \leq A$. Since the messages are transmitted simultaneously in a non-orthogonal fashion, detection at each AU follows that of NOMA, which involves power domain multiplexing at the transmitter and successive interference cancellation (SIC) at each AU \cite{sun2015ergodic,choi2017noma,ali2018downlink}. Based on the above, for all $i  \in \llbracket 1, N \rrbracket$ and $k  \in \llbracket 1, K \rrbracket$, the received signals at the $i$th AU and the $k$th ED are expressed, respectively, as
\begin{equation}
\begin{split}
&y_i = h_i s + n_i \\ 
&z_k = g_ks + w_k,
\end{split}
\end{equation}
where $h_i, g_k \in \mathbb{R}^+$ are the channel gains of the $i$th AU and the $k$th ED, and $n_i$ and $w_k$ are AWGN samples that are $\mathcal{N} \left(0, \sigma_A^2 \right)$ and $\mathcal{N} \left(0, \sigma_E^2 \right)$ distributed, respectively.. For all $i  \in \llbracket 1, N \rrbracket$ and $k  \in \llbracket 1, K \rrbracket$, let $\rho_{A,i} \stackrel{\vartriangle}{=} \frac{h_i^2}{\sigma_A^2}$ and $\rho_{E,k} \stackrel{\vartriangle}{=} \frac{g_k^2}{\sigma_E^2}$ denote the normalized received SNR at the $i$th AU and at the $k$th ED, respectively. Finally, let $\rho_{E} = \sum_{k=1}^K \rho_{E,k}$. In the following, we review all the works and results reported in the literature of secure SISO VLC systems for the cases of single AU multiple AUs.
\subsection{Single AU}
\begin{figure}[t]
	\centering
	\includegraphics[width=1\linewidth]{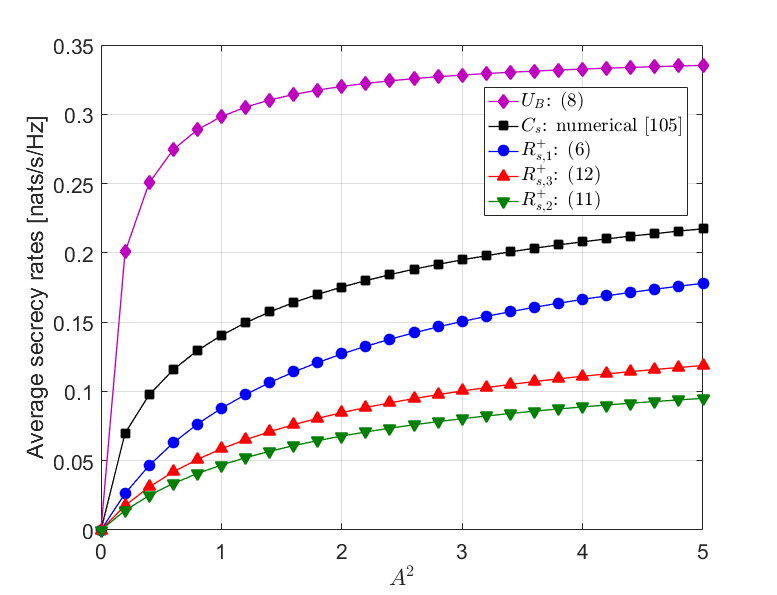}
	\caption{Upper bound $U_B$, secrecy capacity $C_s$ and lower bounds $R_{s,1}^+$, $R_{s,2}^+$ and $R_{s,3}^+$ versus $A^2$. $\sigma_A^2 = \sigma_E^2 = -98.83$ [dBm].}
	\label{fig:Sim3}
\end{figure}

In this subsection, we assume that the transmitter communicates with $N = 1$ AU. In this case, note that if $\rho_{A,1} \leq \rho_{E}$, then the SISO VLC wiretap channel is not degraded and, therefore, the secrecy capacity $C_s = 0$ \cite{ozel2015gaussian}. However, If $\rho_{E} < \rho_{A,1}$, then the SISO VLC wiretap channel is strictly degraded and the secrecy capacity is not null \cite{ozel2015gaussian}. Therefore, we assume to the end of this part that $\rho_{E} < \rho_{A,1}$. In this case, It was shown in \cite{ozel2015gaussian} that the secrecy capacity achieving probability distribution is unique, symmetric and discrete with a finite support set. However, neither the secrecy capacity nor the optimal input probability distribution were derived in closed-form and it remains an open problem. Therefore, several upper and lower bounds were derived in the literature in an attempt to characterize the secrecy capacity of the system. \\ 
\indent The problem of secure SISO VLC systems with a single AU and single ED with perfect CSI was investigated in \cite{mostafa2015physical,wang2018physical,wang2018secrecyCF,arfaoui2018secrecy}. In \cite{mostafa2015physical}, an achievable secrecy rate for the system was proposed and it is given by $R_{s,2}^+$, such that
\begin{equation}
\begin{split}
&R_{s,2} = \frac{1}{2} \log \left( 1 + \frac{2 A^2}{2 \pi e} \rho_{A,1}  \right)  - \frac{\delta}{\sqrt{2 \pi \sigma_E^2}} \exp \left( \frac{-\delta^2}{2 \sigma_E^2} \right) \\ 
&- \left(1 - 2 \mathcal{Q} \left( \frac{\delta}{\sigma_E} + \sqrt{\rho_{E}} A \right) \right) \log \left( \frac{2 \left(\sqrt{\rho_{E}} A + \delta \right)}{\sqrt{2 \pi} \left(1 - 2 \mathcal{Q} \left( \frac{\delta}{\sigma_E} \right) \right)} \right) \\ 
&- \mathcal{Q} \left( \frac{\delta}{\sigma_E} \right) + \frac{1}{2}, 
\end{split}
\end{equation}
where $0 < \delta$ is a free parameter and $\mathcal{Q} \left( \cdot \right)$ is the $\mathcal{Q}$-function. In \cite{wang2018physical,wang2018secrecyCF}, the authors considered the illumination requirement of LEDs and added an average optical power constraint along with the peak-power constraint. Then they derived upper and lower bounds on the secrecy capacity of the system. \\ 
\indent Based on the type of the input distribution $p_s$, i.e., either continuous or discrete over the interval $[-A,A]$, two achievable secrecy rates were proposed in the literature. Specifically, assuming that the transmitted signal $s$ is continuous, an achievable secrecy rate for the system, that was proposed in \cite{arfaoui2018secrecy}, is given by $R_{s,3}^+$, such that 
\begin{equation}
R_{s,3} = \frac{1}{2} \log \left( 1 + \frac{\exp\left(2 h_{u_1} \right)}{2 \pi e} \rho_{A,1}  \right) - \frac{1}{2} \log \left(1 + \sigma_{u_1}^2 \rho_{E} \right),
\end{equation}
where $h_{u_1}$ denotes the differential entropy $u$ and $\sigma_{u_1}^2$ denotes its variance. Several continuous input distributions were proposed in the literature, which aim to maximize the achievable secrecy rate $R_{s,3}$, including the uniform distribution, the truncated Gaussian distribution and the truncated generalized normal (TGN) distribution \cite{arfaoui2018secrecy}. Furthermore, by using discrete probability distributions, another achievable secrecy rate was proposed in \cite{arfaoui2018secrecy}. This achievable secrecy rate is $R_{s,1}^+$ in (6) for the special case when $(M,N,K) = (1,1,1)$.  \\
\indent Fig.~\ref{fig:Sim3} presents the upper bound $U_B$ in (8), the numerical calculation of the secrecy capacity $C_s$ and the lower bounds $R_{s,1}^+$, $R_{s,2}^+$ and $R_{s,3}^+$, obtained through $10^5$ independent Monte-Carlo trials on the location of the AU and the ED, versus $A^2$. The secrecy capacity is obtained numerically by using the same approach invoked in \cite{ozel2015gaussian}, whereas for the lower bounds $R_{s,1}^+$ and $R_{s,3}^+$, the TGN and the TDGN distributions were adopted, respectively. Specifically, the best parameters of the TGN and the TDGN distributions are obtained through brute force (BF) search methods. As shown in this figure, the lower bound $R_{s,1}^+$, the one that corresponds to the discrete input distributions with finite support sets, is the best lower bound, which is not surprising since it is already known that the optimal input distribution for the degraded SISO VLC wiretap channel is discrete with a finite support set \cite{ozel2015gaussian}. \\
\indent When the EDs' CSI is not available at the transmitter, it becomes more challenging to characterize the secrecy capacity and quantify the achievable secrecy rate. To overcome this obstacle, researchers have proposed invoking stochastic geometry approaches \cite{arfaoui2018secrecy,cho2018impact,cho2018physical}. In this context, the locations of the EDs are treated as random variables, and therefore, one may quantify the achievable secrecy rates by averaging over all possible locations. In \cite{arfaoui2018secrecy}, only one randomly located ED is considered, whereas in \cite{cho2018impact,cho2018physical}, multiple randomly located EDs were considered. All receivers, including that of the AU, were assumed to be uniformly distributed within the coverage area and their orientation was assumed to be fixed and facing the transmitter. Specifically, in \cite{arfaoui2018secrecy}, the statistics of the received SNRs $\rho_{A,1}$ and $\rho_{E}$ were derived, which lead to evaluating the secrecy performance of the system in terms of the average secrecy rates $\mathbb{E}\left(U_B \right)$, $\mathbb{E}\left(R_{s,1}^+ \right)$ and $\mathbb{E} \left(R_{s,3}^+ \right)$. In \cite{cho2018impact,cho2018physical}, the PPP process was used to model the number of EDs. The main differences between \cite{cho2018impact,cho2018physical} are as follows. The EDs were assumed to be non-colluding in \cite{cho2018impact} and colluding in \cite{cho2018physical}. In addition, the authors in \cite{cho2018impact} studied the impact of multipath reflections on the secrecy performance, whereas only the LoS component of the channel gains was considered in \cite{cho2018physical}. In both papers, the statistics of the received SNRs were derived, which facilitates evaluating the secrecy performance in terms of the SOP. Finally, the lower bound $R_{s,3}^+$ in conjunction with uniform distribution was adopted as an achievable secrecy rate.
\subsection{Multiple AUs}
The problem of secure MU-SISO VLC channels was investigated in \cite{arafa2018relay,arafa2018relayJ,pan2017secure,zhao2018physical}. The authors in \cite{arafa2018relay,arafa2018relayJ} considered a system comprising a transmitter that intends to send $N=2$ confidential messages to two AUs in the presence of an external ED. In order to enhance the secrecy performance of this broadcast system, the authors investigated two transmission strategies: the direct transmission and the relay-aided transmission. For the direct transmission strategy, superposition coding with uniform signaling was used, whereas in the relay-aided transmission strategy, three secure transmission schemes were studied, which are 
\begin{enumerate}
    \item Cooperative jamming: the relay acts as a jammer and sends artificial noise signals in order to degrade the ED reception. 
    \item Decode-and-forward: the transmitter sends the information bearing signal to the relay, which in turn decodes and then forwards it to the AU.
    \item Amplify and forward: the transmitter sends the information bearing signal to the relay, which in turn amplifies it by infusing more optical power and then forwards it to the AU.
\end{enumerate}
For all these transmission schemes, the authors derived the achievable secrecy regions and studied the performance of each scheme with respect to the geometry of the coverage area and the location of the ED. \\ 
\indent The system model considered  in \cite{pan2017secure,zhao2018physical} consists of a transmitter communicating with $N$ AUs in the presence of a set of EDs, where their number is modeled using a PPP model. Specifically, in \cite{pan2017secure}, only the AU at the room center was supposed to be a legitimate user, while all other users were supposed to be EDs. In addition, all the EDs, including the remaining AUs, were assumed to be non-cooperative, i.e., they do not share the eavesdropped information. In this case, only the ED with the highest received SNR $\rho_{E,max}$ was considered as a potential eavesdropper. Based on this, the statistics of $\rho_{E,max}$ were derived, which yielded a closed-form expression of the SOP. However, since the closed-form expression of the secrecy capacity with the input amplitude constraint is not readily available, the amplitude constraint was ignored and Gaussian signaling was adopted instead. In \cite{zhao2018physical}, a two-user NOMA-SISO VLC system was considered. The transmitter communicates simultaneously with two uniformly distributed AUs within the coverage area. Assuming that the EDs are not colluding and by considering only the ED with the highest SNR, the authors derived a closed-form expression for the SOP for each AU.
\section{Hybrid VLC/RF systems}
\begin{figure}[t] 
\centering
\includegraphics[width=1\linewidth]{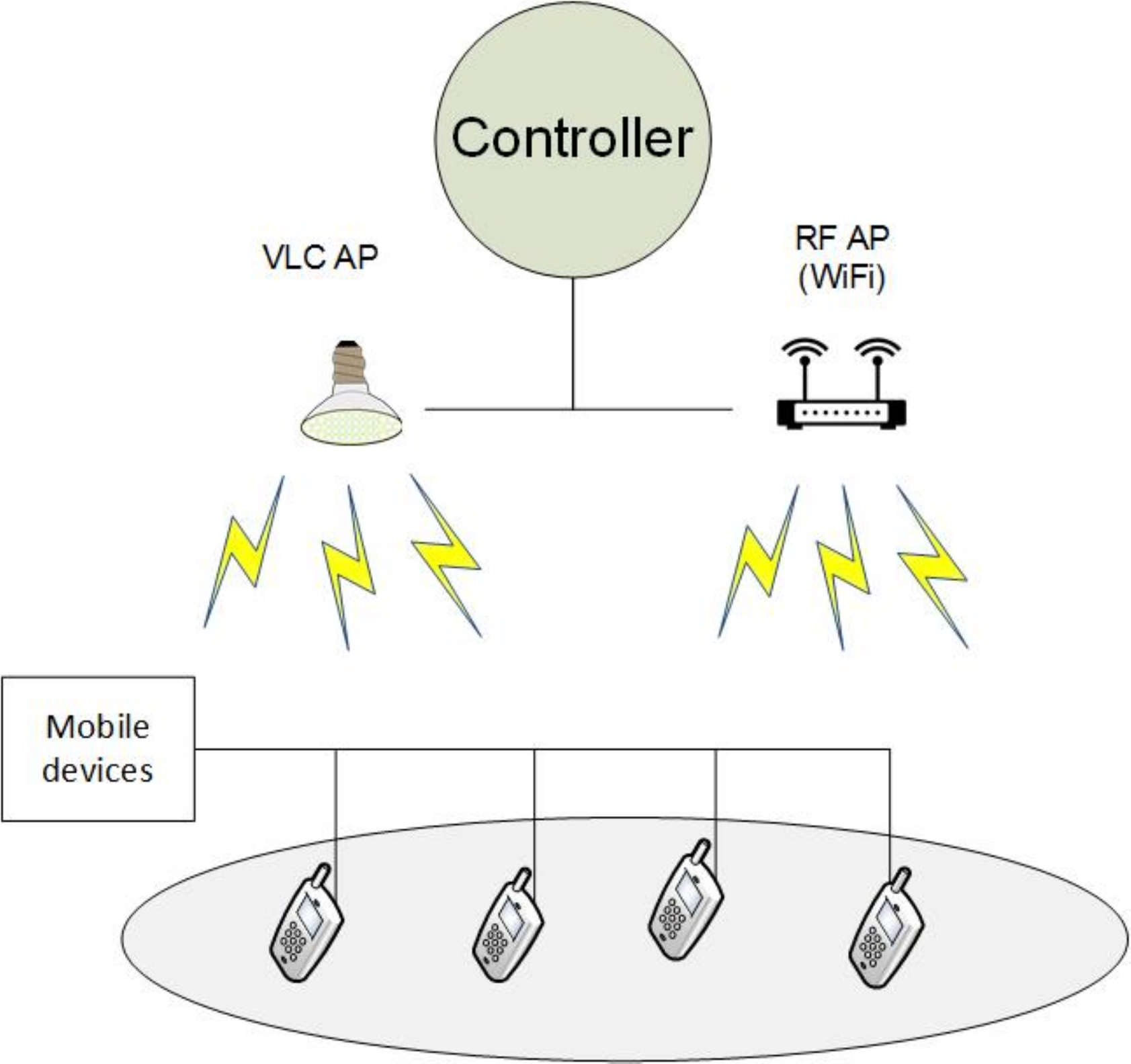}
\caption{Illustration of the heterogeneous LiFi-WiFi system.} 
\label{fig:RFVLC} 
\end{figure}

In indoor environments, the light diffused from LED sources is naturally confined to a small area. In addition, the light beams are susceptible to indoor blockages, such as the human body, which may cause severe fluctuations in the received SNR. Consequently, the hybrid integration of VLC and RF systems was envisioned to significantly improve the user experience, since VLC systems can support very high data rates in specific areas and RF systems can provide greater coverage area to support mobility  \cite{rahaim2011hybrid,wang2015efficient,ArdimasLiFiRF}. In addition, one of the challenges of adopting VLC systems in indoor environments is the VLC uplink transmission mechanism, especially if one relies on the existing indoor infrastructure for using LEDs for illumination and as VLC APs. To overcome these limitations, two alternatives were proposed in the literature. The first alternative is the invention of light-fidelity (LiFi) technology. LiFi is a fully networked optical communication system that includes both uplink and downlink transmissions \cite{soltani2018bidirectional}. The uplink operates over the infrared (IR) spectrum while the downlink operates over the visible light spectrum \cite{soltani2018bidirectional}. The second alternative consists of deploying hybrid wireless fidelity (WiFi)/LiFi. Both technologies are bidirectional and can jointly provide high coverage and high data rates. Fig~\ref{fig:RFVLC} presents an illustration of the hybrid WiFi/LiFi architecture, where the RF AP is a WiFi node and the VLC AP is an array of LEDs. Since hybrid VLC/RF systems encompass VLC components and RF components jointly, security for hybrid VLC/RF systems should be attentively investigated due to the broadcast nature of both communication systems, which is the focus of this section. \\ 
\indent Security for hybrid VLC/RF networks was investigated in \cite{marzban2017beamforming,pan2017secure2,pan2017secrecy,al2019secrecy}. In \cite{marzban2017beamforming}, a downlink hybrid RF/VLC wiretap system was considered, where one RF AP, equipped with multiple antennas, and one VLC transmitter, equipped with multiple LEDs, communicate jointly with one AU in the presence of one ED. The two receivers were assumed to have the multi-homing capability, i.e., they can aggregate the information from the RF and VLC transmitters. The authors employed beamforming as a transmission strategy and then optimized over the beamforming vectors and the transmit powers at both the VLC and RF AP. Then, they formulated the PLS scheme in order to minimize the total consumed power while achieving target RF and VLC information rates at the AU and maintaining zero information rate at the ED. In \cite{pan2017secrecy,pan2017secure2}, a bidirectional hybrid RF/VLC wiretap system was considered, where one VLC transmitter, equipped with one AP, transmits confidential messages to a randomly located AU in the presence of one randomly located ED. In addition, it was assumed that the two receivers harvest energy from the light intensity. In addition, the AU communicates with an RF AP and this uplink transmission of the information is performed over an RF link that suffers from Rician fading, while the ED tries to acquire the transmitted information. The authors derived the secrecy outage probability using stochastic geometry and evaluated the secrecy performance of the system with respect to the geometry of the coverage area. Recently, the authors in \cite{al2019secrecy} investigated the secrecy performance of a hybrid VLC/RF system equipped with decode-and-forward relaying, where they developed PLS algorithms based on ZF beamforming techniques that mitigate eavesdropping in both RF and VLC networks. 
\section{Future Research Directions}
Although many papers have been published on various aspects of VLC systems, we believe there are still many open problems that need to be addressed to bring the potential of VLC-related technologies to their full potential. Since this paper is concerned with security, we dedicate the rest of this section to highlighting a number of open research problems related to security for VLC.
\subsection{Input Signaling Schemes}
The input signaling schemes present a fundamental feature in designing wireless communication systems. In fact, any transmission scheme and the resulting system performance depend on the type of adopted input signaling scheme. In general, each input signaling scheme consists of two main parts, namely, the input probability distribution and the precoding scheme, also known as the transmission strategy. The input probability distribution defines the probability distribution according to which the signals are transmitted. This can be either continuous or discrete. The precoding scheme defines the form or the shape of the transmitted signal, and it includes the modulation scheme as well. Among the well known precoding schemes for the case of a single AU, one can cite beamforming, artificial noise based beamforming, OFDM and spatial modulation \cite{renzo2014spatial,di2011spatial}, etc. On the other hand, for the case of multiple AUs, one can cite linear precoding, such as GSVD-based precoding and ZF precoding, orthogonal multiple acces (OMA) such as orthogonal frequency division multiple access (OFDMA) \cite{miridakis2013survey}, NOMA \cite{ding2017survey,dai2018survey} and spatial modulation. As shown in previous sections, the input signaling scheme has a high impact on the secrecy performance of VLC systems. However, the optimal input signaling scheme for secure VLC systems is still an open problem. \\ 
\indent Recall that VLC systems impose a peak-power constraint, i.e. amplitude constraint, on the channel input, which is fundamentally different from the average power constraint. In fact, when average power constraints are imposed, it was shown that Gaussian signaling schemes are optimal \cite{khisti2010secure2}. However, from an information theoretic point of view, finding the optimal input signaling schemes that achieve the secrecy capacity of a Gaussian wiretap channel under an amplitude constraint is still an open problem. This is attributed to the fact that, when input distributions of unbounded support like Gaussian inputs are not permissible, the optimal input distribution is either unknown or only known to be discrete for the special cases of a degraded SISO wiretap channel \cite{ozel2015gaussian} and a MISO wiretap channel \cite{rezki2017secret}. Given that VLC falls into this category since amplitude constraints must be satisfied, the optimal signaling scheme for secure VLC systems is still an open problem.
\subsection{PLS for Indoor VLC Systems: What is Missing?}
\indent The basic PLS techniques employed to secure VLC systems that are reported in the literature are summarized in Table II. By having a closer look at this table, one can note that there is a good number of problems related to secure indoor VLC systems that have not been resolved or even considered. A list of these problems is detailed on the following. 
\begin{figure*}[t] 
\begin{center}
\renewcommand{\arraystretch}{1.2} 
\setlength{\tabcolsep}{0.35cm} 
\begin{tabular}{| l | l | l | c | c | }
	\multicolumn{5}{c}{Table II: PLS techniques for secure indoor VLC systems.} \\
   \cline{4-5} 
  \multicolumn{3}{c|}{} & Continuous signaling scheme & Discrete signaling scheme \\ 
   \hline 
   \multirow{4}{*}{MIMO} & \multirow{2}{*}{Single AU} & Single ED & \cite{arfaoui2017mimo}  & \cite{arfaoui2018TWC2}  \\ 
   \cline{3-5} 
   & & Multiple EDs  & $\varnothing$ & \cite{le2014secured}  \\ 
   \cline{2-5} 
   & \multirow{2}{*}{Multiple AUs} & Single ED & \multicolumn{2}{c|}{$\varnothing$} \\ 
   \cline{3-5}
   & & Multiple EDs  & \cite{chen2017physical}  & $\varnothing$  \\
   \hline
   \multirow{4}{*}{MISO} & \multirow{2}{*}{Single AU} & Single ED & \cite{mostafa2014physical,mostafa2015physical,mostafa2016optimal,arfaoui2016input,mostafa2014securing,zaid2015improved,chosecuring,chosecuringJ,arfaoui2016secrecy}  & \cite{arfaoui2017discrete,shen2016secrecy,arfaoui2018TCOM,wang2018optical,wang2018secrecy,li2018secrecy}  \\ 
   \cline{3-5} 
   & & Multiple EDs  & \cite{cho2018securing,arfaoui2018enhancing,chow2015secure} & \cite{arfaoui2018discrete}  \\ 
   \cline{2-5} 
   & \multirow{2}{*}{Multiple AUs} & Without EDs & \cite{mostafa2017linear,pham2015max,pham2017multi,arfaoui2018TWC1}  & $\varnothing$ \\ 
   \cline{3-5} 
   & & Single ED & \cite{pham2016secrecy,pham2017secrecy,pham2018artificial,pham2018artificialJ}  & $\varnothing$ \\ 
   \cline{3-5} 
   & & Multiple EDs  & \cite{yin2018physical} & $\varnothing$  \\ 
   \hline
   \multirow{4}{*}{SISO} & \multirow{2}{*}{Single AU} & Single ED & \cite{mostafa2015physical,wang2018physical,wang2018secrecyCF,arfaoui2018secrecy}  & \cite{arfaoui2018secrecy}  \\ 
   \cline{3-5} 
   & & Multiple EDs  & \cite{cho2018impact,cho2018physical} & $\varnothing$  \\ 
   \cline{2-5} 
   & \multirow{2}{*}{Multiple AUs} & Single ED & \cite{arafa2018relay,arafa2018relayJ} & $\varnothing$  \\ 
   \cline{3-5} 
   & & Multiple EDs  & \cite{pan2017secure,zhao2018physical}& $\varnothing$  \\ 
   \hline
    \multirow{4}{*}{Hybrid RF/VLC} & \multirow{2}{*}{Single AU} & Single ED & \cite{marzban2017beamforming,pan2017secure2,pan2017secrecy,al2019secrecy}  & $\varnothing$ \\ 
   \cline{3-5} 
   & & Multiple EDs  & \multicolumn{2}{c|}{}    \\ 
   \cline{2-3} 
   & \multirow{2}{*}{Multiple AUs} & Single ED &  \multicolumn{2}{c|}{$\varnothing$}   \\ 
   \cline{3-3} 
   & & Multiple EDs  &  \multicolumn{2}{c|}{}  \\ 
   \hline
\end{tabular} 
\end{center}
\end{figure*} 
\begin{enumerate}
\item For the MU-MISO VLC broadcast channel considered in subsection \ref{MultipleAUs}, all previous works used only continuous input distributions. However, adopting continuous input signaling is unrealistic since digital data streams should be transmitted. In addition, discrete input signaling has been shown to be optimal for the two-user IM-DD discrete memoryless free space optical broadcast channels with non-negativity, peak and average intensity constraints at the transmitter \cite{soltani2018capacity}. Although it is not straightforward to derive optimal precoding schemes for secure communications in MU-MISO VLC broadcast channels with an arbitrary number of AUs using discrete distributions, there is a good chance that the secrecy performance of these systems may be enhanced by adopting this class of input distributions.
\item In the same context, note that the majority of works in the literature that investigated the secrecy performance of MU-MISO VLC broadcast channels assumed that there is at least one passive ED that can exist within the coverage area. Such an assumption may be neither valid nor realistic in some practical scenarios. In fact, from a security point of view, one can not predict or estimate the number of potential EDs existing in the same coverage area that are capable of extracting confidential information from legitimate users. Therefore, investigating the secrecy performance of MU-MISO VLC systems serving a set of AUs in the presence of a set of passive EDs may be interesting, since it would be a generalization of the results derived in the existing research works. 
\item The secrecy performance of the MU-MIMO VLC broadcast channel adopted in has not been investigated in the literature. In a typical MU-MIMO VLC broadcast channel, the transmitter is equipped with multiple APs and serves multiple AUs, each equipped with multiple PDs, in the presence of a set of colluding EDs with imperfect CSI. This model is not only a generalized model that encompasses all the VLC systems adopted in the literature but also the most realistic one \cite{yang2013optimal,mukherjee2009user}. Therefore, investigating the secrecy performance of this system is also an important contribution to the topic of secure VLC systems. 
\end{enumerate}
\subsection{Incorporating Realistic and Measurements Based VLC Channel Models} 
\begin{figure}[t!]
	\centering
	\begin{subfigure}[b]{0.72\columnwidth}
		\centering
		\includegraphics[width=1\columnwidth,draft=false]{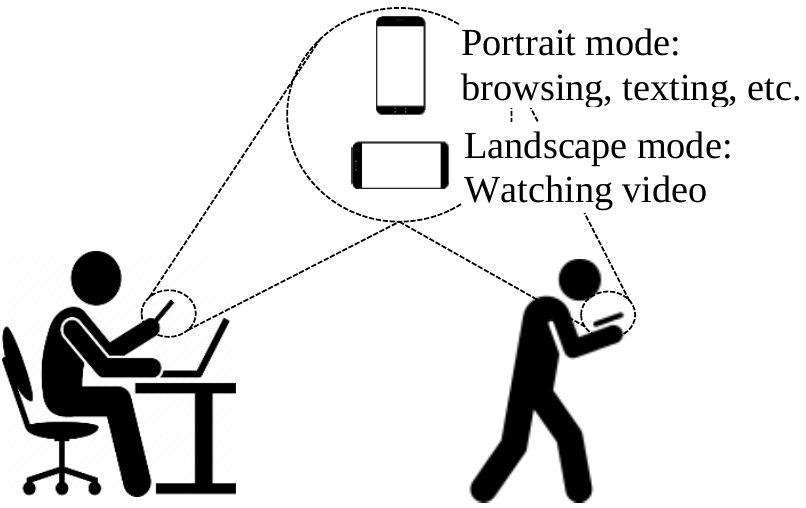}
	    \caption{Measurement setup.}
		\label{sub1:RWP}
	\end{subfigure}~
	\begin{subfigure}[b]{0.28\columnwidth}
		\centering
		\includegraphics[width=1\columnwidth,draft=false]{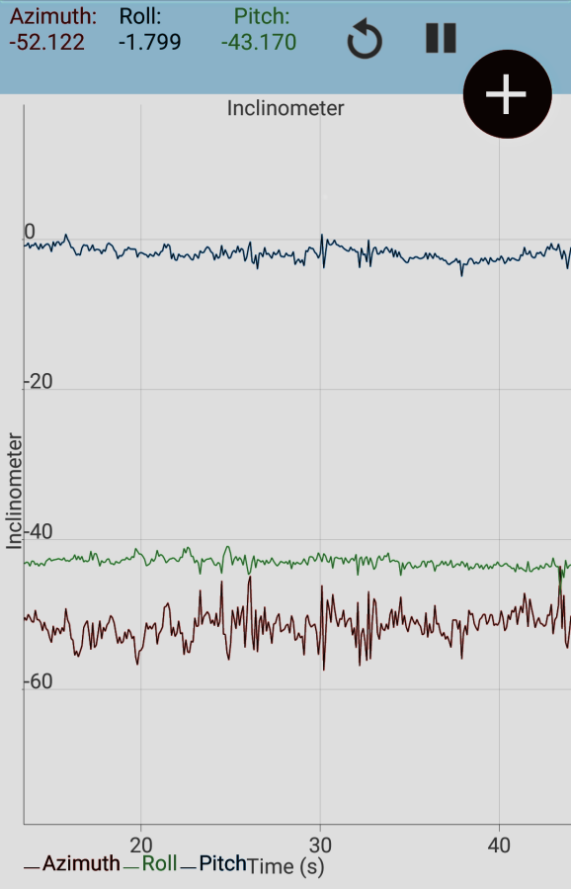}
		\caption{Sensor App.}
		\label{sub2:RWP}
	\end{subfigure}
	\\
	\begin{subfigure}[b]{0.5\columnwidth}
		\centering
		\includegraphics[width=1\columnwidth,draft=false]{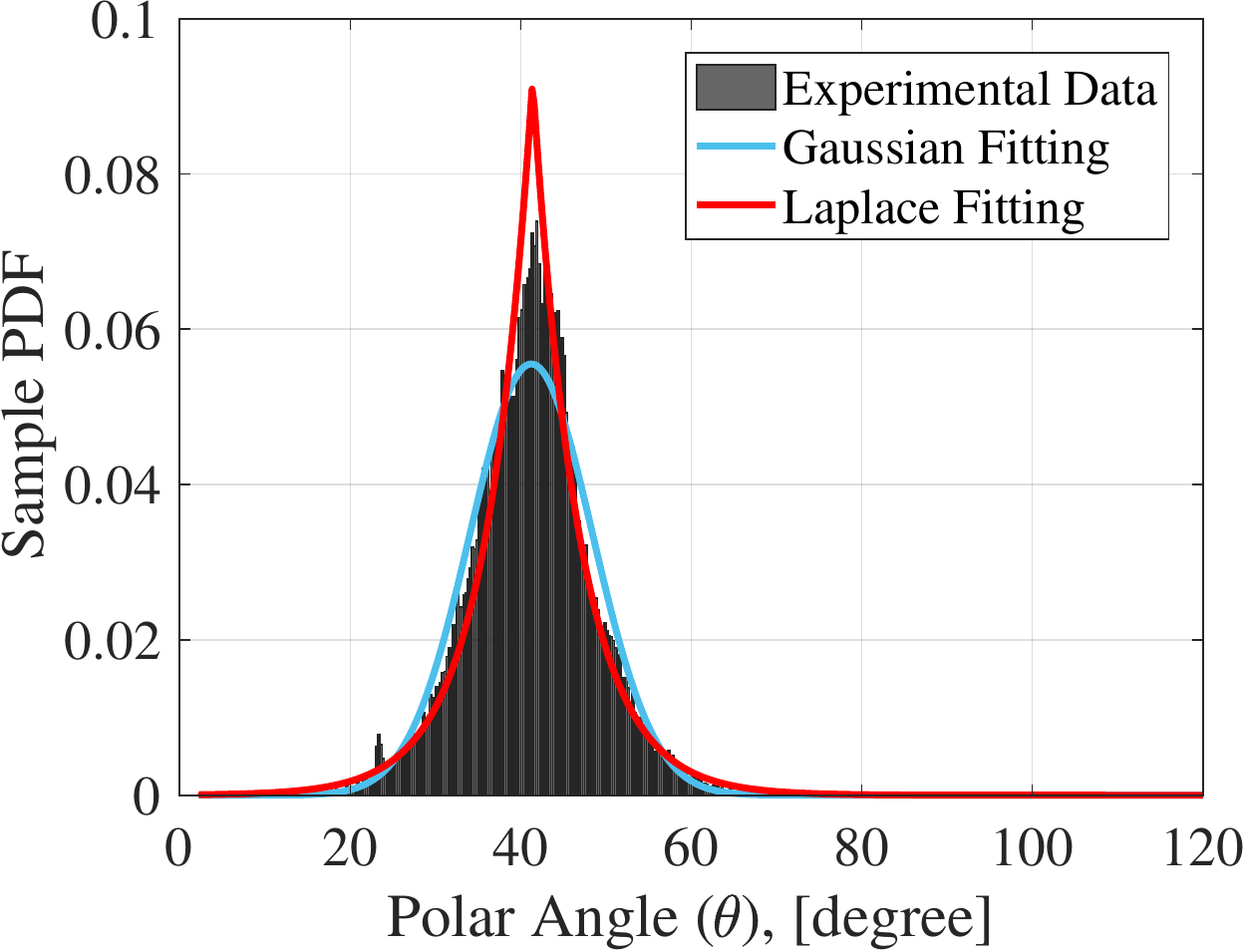}
		\caption{Sitting activities \cite{soltani2018modeling}.}
		\label{sub3:RWP}
	\end{subfigure}
	\begin{subfigure}[b]{0.5\columnwidth}
		\centering
		\includegraphics[width=1\columnwidth,draft=false]{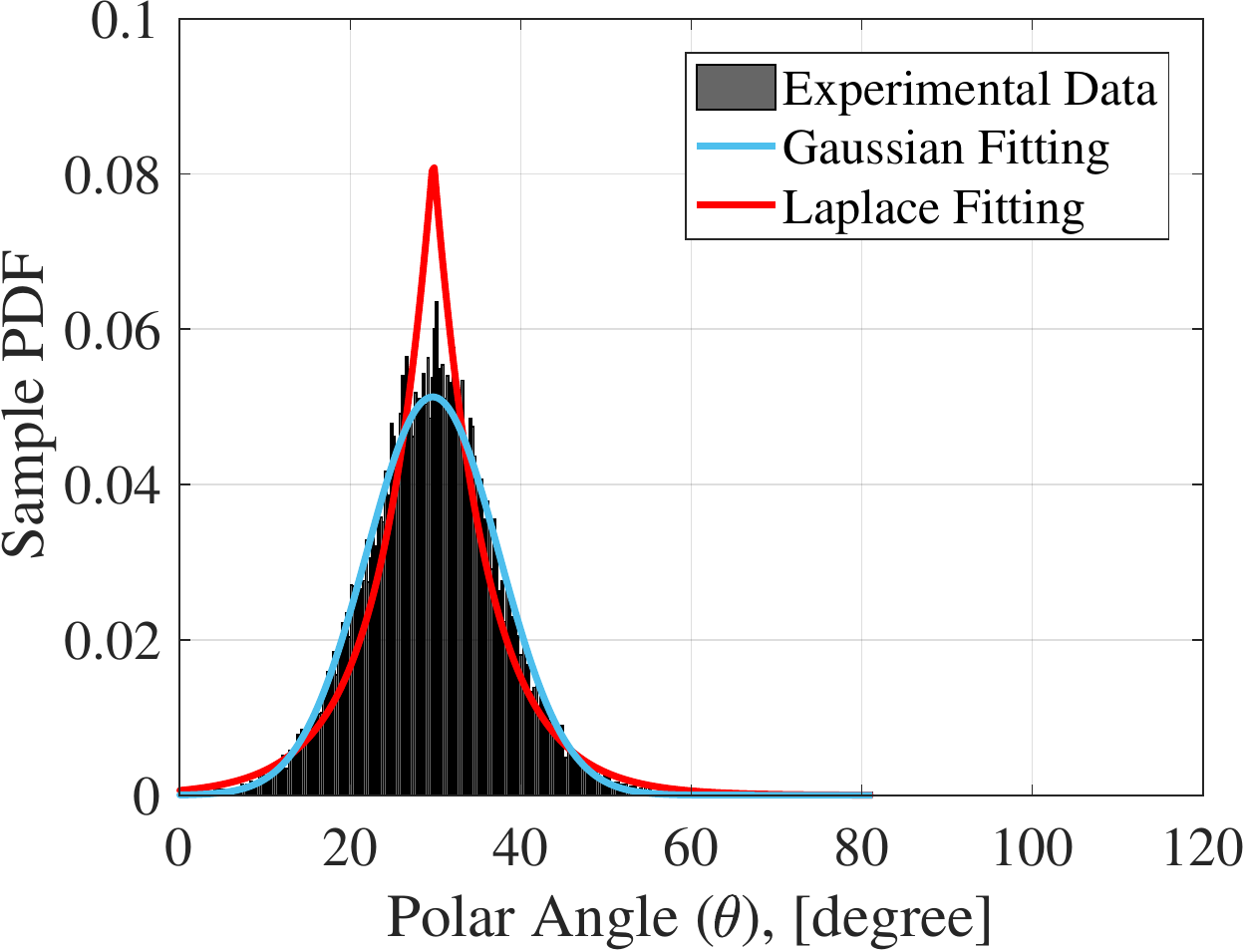}
		\caption{Walking activities \cite{soltani2018modeling}.}
		\label{sub4:RWP}
	\end{subfigure}
	\caption{Illustration of experimental setup for both sitting and walking activities.}
	\label{Orientation_measurements}
\end{figure}

All of the studies on secure VLC systems reported in the literature employed the VLC channel model presented in section \ref{SY1}. In this model, it usually assumed that the receiver in an indoor VLC system is either stationary or uniformly distributed within the area of interest and that its orientation is constant (vertically upward and fixed). However, the presence of mobility and device orientation as well as link blockage, which are inherent features of wireless networks, require more realistic and non uniform models. In addition, none of the previous studies on secure VLC systems have considered the actual statistics of the receiver's mobility and orientation. This is attributed to the simplicity of the adopted model and also to the lack of proper channel models for the mobility and orientation of VLC devices in indoor scenarios. Nevertheless, several measurement-based channel models for VLC system were derived recently for various environments, such as indoor VLC systems \cite{miramirkhani2017mobile,miramirkhani2015channel,purwita2018WCNC,Zhihong_VTCfall_2018,soltani2018modeling,mohammad2018optical,purwita2018terminal}, VLC-based vehicular communication \cite{elamassie2018effect} and underwater VLC systems \cite{elamassie2018performance,miramirkhani2018visible}.  \\
\indent In general, a receiver in a VLC system can be mobile and can have a random orientation. From section \ref{SY1}, the channel gain of an indoor VLC receiver is expressed as 
\begin{equation}
h= C  \cos(\theta)^m \frac{ \cos(\psi)}{d^2} \text{rect} \left( \frac{\psi}{\Psi} \right), 
\end{equation}
where $C = \dfrac{\eta \, R \, T \, (m+1)}{2\,\pi} \frac{n_c^2 A_{g}}{\sin(\Psi)^2}$. The user's mobility will induce randomness in the distance $d$, the angle of transmission $\theta$ as well as the angle of incidence $\psi$, whereas the random orientation of the user's device will induce randomness only in the angle of incidence $\psi$. Recently, the authors in \cite{purwita2018WCNC,Zhihong_VTCfall_2018,soltani2018modeling,mohammad2018optical} derived measurement-based VLC channel models for indoor VLC receivers with random orientation, where they showed that the incidence angle $\psi$ follows a truncated Laplace distribution if the receiver is stationary and a truncated Gaussian distribution if the receiver is mobile. The experimental measurements were obtained for both sitting and walking activities, in portrait and landscape modes as shown in Fig.~\ref{Orientation_measurements}. The impact of device orientation on LoS link availability is assessed in \cite{MDSWCNC19}. In \cite{soltani2018impact}, the authors used the Laplace distribution to derive the pdf of SNR and BER of indoor VLC systems analytically. It was shown that the random orientation of a device can influence the user's performance, which depends on its location. \\
\indent Considering the receiver's mobility, several mobility models have been proposed in the literature to model the distribution of mobile users in indoor scenarios. The most commonly used mobility model in indoor systems is the random waypoint (RWP) mobility model \cite{bettstetter2003node,hyytia2005random,hyytia2006spatial,ali2011performance,MDSHandover}. In this stochastic model, each user of the network uniformly chooses a random destination point (“waypoint”) in a given deployment area. A user moves to this destination with a velocity $v$ chosen uniformly in the interval $[v_{\min}, v_{\max}]$. When it reaches the destination, it remains static for a random pause time and then starts moving again according to the same rule. Recently, the authors in \cite{gupta2018statistics,arfaoui2019SNR} derived the channel statistics for indoor mobile VLC receivers using the RWP mobility model, where the receiver's orientation is assumed to be fixed in \cite{gupta2018statistics} and uniformly distributed in \cite{arfaoui2019SNR}. In addition, motivated by the results of \cite{gupta2018statistics}, the authors in \cite{arfaoui2019effect} investigated the secrecy performance of the strongest AU in a SISO NOMA VLC system in the presence of multiple mobile and colluding EDs, where the receivers' orientation was assumed to be fixed and their mobility was characterized using the RWP mobility model. Based on this, the channel statistics and the SOP of the strongest AU were derived using stochastic geometry. \\ 
\indent Based on the above discussion, re-investigating the secrecy performance of VLC systems by considering the aforementioned measurement-based channel models and mobility models presents an interesting future research direction, since it helps to develop adequate PLS techniques for VLC systems in practical and real-life scenarios. Furthermore, in order to analyze the performance of mobile users more realistically, an orientation-based RWP (ORWP) mobility model was first proposed in \cite{soltani2018modeling} and then extended in \cite{mohammad2018optical}. The ORWP model provides a more realistic framework for performance evaluation of mobile users in VLC systems by incorporating the orientation of the device and the mobility of the user. Therefore, studying the secrecy performance for mobile users according to the ORWP mobility model can be another future research topic.  \\ 
\indent The VLC channel is also susceptible to link blockage. Due to the inherent nature of the VLC channel, the link between a pair of receiver and transmitter can be interrupted by obstructions such as a human body or other similar objects. However, it is noted that the main cause of link blockage is the mobile users in the indoor environment. The blockers can be modeled either as cylindrical objects \cite{6354257} or rectangular prisms \cite{8471819}. 
When the communication link is interrupted by blockers, adding extra power cannot compensate for the data loss. Solutions to simplify the effect of link blockage were introduced in \cite{mohammad2018optical,ImanICCW19,ChengICCW19}. A multi-directional receiver (MDR) for which PDs are located at different sides of a smartphone is introduced in \cite{mohammad2018optical,ImanICCW19} and has been assessed in the presence of link blockage, user mobility as well as device random orientation. It is denoted that the MDR structure outperforms the conventional configuration for which all PDs are placed on one side of a smartphones, for example on the screen side. In \cite{ChengICCW19}, the author proposed an omnidirectional receiver in which PDs are embedded on all sides of a smartphone. This configuration is a robust scheme against blockage. Its superior performance is compared to a single-PD and two-PD configurations. Accordingly, the secrecy performance of the VLC system can be evaluated in the presence of blockers with different densities in an indoor environment. Moreover, the MDR configuration and omnidirectional receiver are the other two interesting future perspectives that should be studied from a secrecy performance aspect. 
\begin{figure*}[t] 
\centering
\includegraphics[width=1\linewidth]{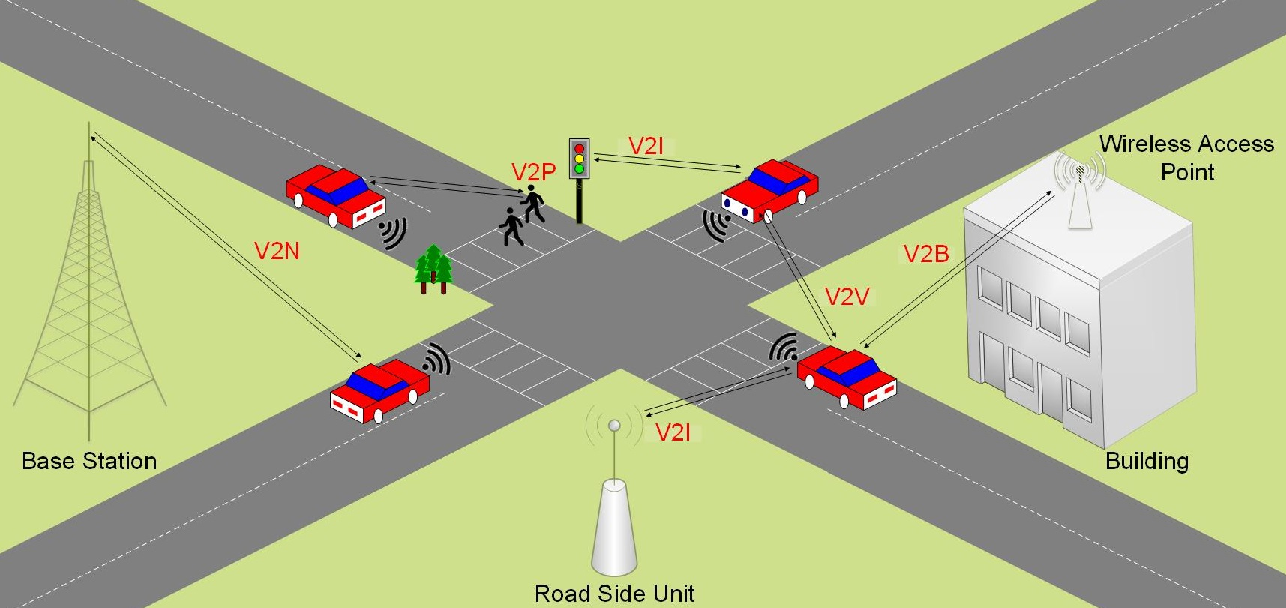}
\caption{Illustration of V2X communication.} 
\label{fig:V2X} 
\end{figure*}
\subsection{Security for Outdoor VLC systems}
Outdoor VLC applications are less explored when compared to their indoor counterparts. This is mainly due to two facts: 1) the dual use of light-emitting diodes is not always practical in an outdoor VLC environment, and 2) the level of interference and noise is considerably higher in outdoor environments. Nevertheless, several outdoor VLC applications have been identified in the literature. The adoption of VLC in outdoor applications was reviewed in \cite{cuailean2017current}, where the authors revealed the issues that arise in the outdoor usage of VLC, identified emerging challenges and proposed future research directions. In this context, VLC outdoor applications include, but are not limited to, \cite{ndjiongue2018overview},
\begin{enumerate}
\item Vehicle-to-everything (V2X) communication. 
\item Building-to-building (B2B) communication. 
\end{enumerate}
V2X communications is a new concept that uses the latest generation of information and communication technology that connect vehicles to everything. In its turn, and as shown in Fig.~\ref{fig:V2X}, V2X communication includes \cite{wang2019survey}
\begin{enumerate}
\item Vehicle to building (V2B) communication.
\item Vehicle to infrastructure (V2I) communication, also known as road to vehicle (R2V) communication.
\item Vehicle to network (V2N), also known as vehicle-to-cloud (V2C) communications.
\item Vehicle to pedestrian (V2P) communication.
\item Vehicle to vehicle (V2V) communication. 
\end{enumerate}
\begin{figure}[t] 
\centering
\includegraphics[width=1\linewidth]{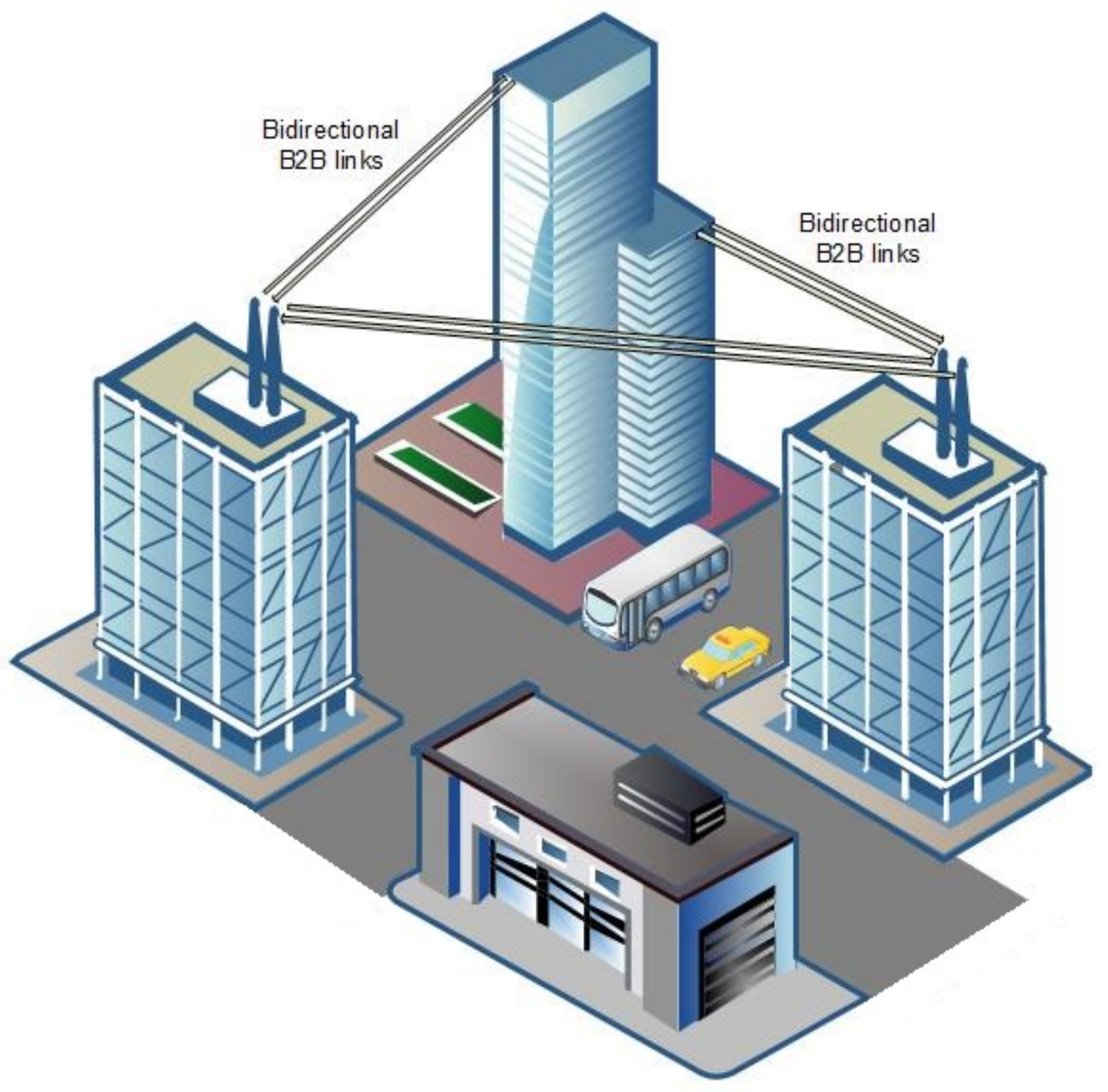}
\caption{Illustration of bidirectional B2B communication.} 
\label{fig:B2B} 
\end{figure}
The V2X technology links the various elements of transportation, such as pedestrians, vehicles, roads, and cloud environments. This leads to the building of an intelligent transport system and promoting the development of new modes and new forms of automobiles and transportation services by gathering more information and promote the innovation and application of automated driving technology \cite{chen2017vehicle}. In addition, V2X communication is of great significance for improving traffic efficiency, saving resources, reducing accidents, and improving traffic management \cite{chen2017vehicle,V2X} \\
\indent On the other hand, B2B communications define all the communications technologies that enable connection between buildings, as shown in Fig~\ref{fig:B2B}. Both RF and FSO (including the visible light and the IR spectra) communications is used for this type of mission \cite{khalighi2014survey}. Therefore, VLC may be used to connect buildings situated within a reasonable distance from each other, such as campuses, bank buildings, and headquarters to provide access to information, data and media. \\
\indent For outdoor applications of VLC, security issues arise naturally for VLC-based V2X communication and VLC-based B2B communication. Specifically, for the case of V2X communications, vehicles commonly use RF, FSO or VLC wireless communication techniques for both V2V and V2I communications. Manufacturers, regulators and the public are understandably concerned about large-scale systems failure or malicious attacks via these wireless vehicular networks. In this context, only a few papers have recently investigated the security problems in V2X communications. In \cite{muhammad2018survey}, common V2X threats as well as security issues and security requirements of V2X in cellular network were reviewed, where the authors discussed existing V2X authentication solutions proposed in the literature. In \cite{rowan2017securing}, the authors presented an implementation and evaluation of the use of ultrasonic audio and image camera visual light side-channels for secure V2V communications. The constraints imposed by these side-channels necessitates the development of new schemes for small throughput, secure and attributable exchange of session key information between vehicles. Therefore, the authors used Blockchain as a V2V message transport, since it provides a secure, verifiable, shared, open and distributed ledger. In \cite{wang2018secure}, the authors studied the secrecy performance of vehicular heterogeneous networks from a PLS point of view. In fact, the considered network model contains an ED, and the security problem was formulated using stochastic geometry. Finally, a secure cooperative communication scheme was proposed to enhance the secrecy performance of the system. Except \cite{wang2018secure}, none of the works reported in the literature has studied the potential of PLS in providing secure V2X communications, for both RF and VLC wireless networks. \\ 
\indent Based on the above, improving the security of VLC-based V2X communications and VLC-based B2B communications from a PLS point of view can be considered as a potential future work. However, some fundamental features should be clearly defined and deeply investigated before performing security analysis. These features include: 
\begin{itemize}
\item The framework and the network configuration, such as distances, dimensions, transmit light sources, optical receivers, etc. 
\item Realistic and measurement-based channel models should be derived. These channel models should encompass the effect of interference as there are fewer physical barriers in outdoor environments (unlike the indoor environment).
\item The effect of sunlight during the day, outdoor illumination during the night, and weather conditions should be studied. It has been shown that solar irradiance does not prevent high speed VLC communication \cite{islim2018impact}. However, careful considerations and system design are required to minimize the effect of those sources of impairment on the performance of VLC systems.
\end{itemize}
\section{Concluding Remarks}
We provided in this paper a comprehensive and comparative review of all PLS techniques reported in the literature that aim to enhance the security of VLC systems. The reported techniques cover both information theoretic and signal processing aspects of VLC systems. Different types of VLC systems were considered, including SISO, MISO and MIMO VLC systems, as well as hybrid RF/VLC systems. In addition, we considered the impact of various VLC features on the secrecy performance, including the input signaling schemes, the geometry and parameters of the network, the number of legitimate receivers and eavesdroppers, and the CSI availability at the transmitting nodes. We also listed a number of open research problems that have great potential for advancing the state-of-the-art of security for VLC systems. What has been accomplished so far in the totality of the research works on security for VLC systems, albeit being fundamental and original, it serves as a starting point for developing realistic PLS techniques tailored to real-world settings in an effort to bring the deployment of VLC-based systems (such as LiFi) closer than ever. 
\section*{Acknowledgment}
M. A. Arfaoui and C. Assi acknowledge the financial support from Concordia University and FQRNT. M. D. Soltani acknowledges the School of Engineering for providing financial support. A. Ghrayeb is supported in part by Qatar National Research Fund under NPRP Grant NPRP8-052-2-029 and in part by FQRNT. H. Haas acknowledges the financial support from the Wolfson Foundation and Royal Society. He also gratefully acknowledges financial support by the Engineering and Physical Sciences Research Council (EPSRC) under an Established Career Fellowship grant EP/R007101/1 and grant EP/L020009/1 (TOUCAN).
\bibliographystyle{IEEEtran}
\bibliography{main.bbl}
\end{document}